\documentclass[aps,prc,twocolumn,showpacs,superscriptaddress]{revtex4}

\usepackage{amsmath}
\usepackage{graphicx,color}

\newcommand{\ba}{\begin{eqnarray}}
\newcommand{\ea}{\end{eqnarray}}

\def\lsim{\,\raise0.3ex\hbox{$<$\kern-0.75em\raise-1.1ex\hbox{$\sim$}}\,}
\def\gsim{\,\raise0.3ex\hbox{$>$\kern-0.75em\raise-1.1ex\hbox{$\sim$}}\,}

\begin{document}

\title{Linear Boltzmann transport for Jet Propagation in the Quark-Gluon Plasma: Elastic Processes and Medium Recoil}


\author{Yayun He}
\affiliation{Key Laboratory of Quark and Lepton Physics (MOE) and Institute of Particle Physics, Central China Normal University, Wuhan 430079, China}
\author{Tan Luo}
\affiliation{Key Laboratory of Quark and Lepton Physics (MOE) and Institute of Particle Physics, Central China Normal University, Wuhan 430079, China}
\author{Xin-Nian Wang}
\affiliation{Key Laboratory of Quark and Lepton Physics (MOE) and Institute of Particle Physics, Central China Normal University, Wuhan 430079, China}
\affiliation{Nuclear Science Division Mailstop 70R0319,  Lawrence Berkeley National Laboratory, Berkeley, CA 94740}
\author{Yan Zhu}
\address{Departamento de F\'{i}sica de Part\'{i}culas and
IGFAE, Universidade de Santiago de Compostela,
E-15706 Santiago de Compostela, Galicia, Spain}

\begin{abstract}
A Linear Boltzmann Transport model within perturbative QCD is developed for the study of parton propagation inside the quark-gluon plasma. Both leading partons and thermal recoil  partons are tracked so that one can also study jet-induced medium excitations. In this study, we implement the complete set of elastic parton scattering processes and investigate elastic parton energy loss, transverse momentum broadening and their nontrivial energy and length dependence. We further investigate medium modifications of the jet shape and fragmentation functions of reconstructed jets. Contributions from thermal recoil  partons are found to have significant influences on jet shape, fragmentation functions and angular distribution of reconstructed jets.
\end{abstract}


\maketitle

\section{Introduction}

When an energetic quark or gluon propagates through a dense partonic system such as the quark-gluon plasma (QGP), it interacts with the hot medium through multiple scattering and induced gluon bremsstrahlung, leading to transverse momentum broadening and energy loss of the propagating parton. As a consequence, spectra of large transverse momentum hadrons and jets in high-energy heavy-ion collisions should be suppressed if a dense partonic matter is formed in the early stage of the collisions. This phenomenon is called jet quenching and has been proposed as a powerful tool for the study of properties of the QGP in high-energy heavy-ion collisions \cite{Bjorken:1982tu,Gyulassy:1990ye,Wang:1991xy}. It has been observed in experiments at both the Relativistic Heavy-ion Collider (RHIC) \cite{Adams:2005dq,Adcox:2004mh} and the Large Hadron Collider (LHC) \cite{Muller:2012zq} as evidence for the formation of strongly interacting QGP in high-energy heavy-ion collisions. Investigators observed not only the suppression of single inclusive hadron spectra at large transverse momentum \cite{Adcox:2001jp,Adler:2002xw,Aamodt:2010jd,CMS:2012aa} but also back-to-back dihadron \cite{stardihadron} and gamma-hadron correlations \cite{Adare:2009vd,Abelev:2009gu}.  The jet quenching phenomenon becomes even more dramatic in heavy-ion collisions at LHC when one observes the onset of dijets and gamma jets with large asymmetries in transverse energy \cite{Aad:2010bu,Chatrchyan:2011sx,Chatrchyan:2012gt}. These patterns of jet quenching can all be understood quantitatively within the picture of multiple scattering and parton energy loss \cite{Wang:2004dn}-\cite{Dai:2012am}.

Within the picture of perturbative QCD (pQCD), the interaction between an energetic parton and the medium is dominated by small-angle scattering and induced gluon radiation. The transverse momentum broadening and energy loss of the propagating parton depend on the transverse momentum transfer from the medium during each scattering which in turn encodes the interaction among medium partons at the scale as probed by the energetic parton. The jet transport parameter defined as the averaged transverse momentum broadening squared per unit length of propagation represents a fundamental property of the medium \cite{Gyulassy:1993hr}-\cite{Arnold:2002ja}.  Recent phenomenological studies of single-hadron suppression in high-energy heavy-ion collisions within collinear factorized pQCD models give extracted values of the transport parameter $\hat q\approx 1.2\pm 0.3$ GeV$^2$/fm and $\hat q\approx 1.9\pm 0.7$ GeV$^2$/fm at an initial time $\tau_0=0.6$ fm/$c$ for a 10 GeV quark at the center of most central Au+Au collisions at $\sqrt{s}=200$ GeV and Pb+Pb collisions at $\sqrt{s}=2.76$ TeV, respectively \cite{Burke:2013yra}. These values are about two orders of magnitude higher than in large cold nuclei \cite{Wang:2009qb,Chang:2014fba}.

Given such large values of the jet transport parameter, the energy lost to the medium by a propagating parton is also very large. Dissipation of this large amount of lost energy in the medium can lead to jet-induced medium excitations such as supersonic waves or Mach cones \cite{CasalderreySolana:2004qm,Stoecker:2004qu}. Study of such jet-induced medium excitations can shed light on the transport properties of the bulk medium which complements that obtained from analyses of anisotropic flows due to collective expansion. Several types of models can be employed to study jet-induced medium excitations such as hydrodynamical models \cite{Chaudhuri:2005vc,Betz:2008ka,Neufeld:2009ep,Qin:2009uh}, string theory \cite{Chesler:2007sv,Gubser:2007ga} and parton transport models \cite{Bouras:2012mh}. The Linear Boltzmann Transport (LBT) model \cite{Li:2010ts,Wang:2010yz,Wang:2013cia,Wang:2014hla} combines a kinetic description of parton propagation and hydrodynamic description of the underlying medium evolution. Since it also keeps track of thermal recoil partons from each scattering and their further propagation in the medium,  the LBT model can be used to study both jet transport in medium and jet-induced medium excitations.

Jet-induced medium excitations in the form of thermal recoil partons and their further propagation are also important for the study of jet quenching through reconstructed jets and the corresponding jet profiles. Reconstructed jets defined through a jet-finding algorithm \cite{Cacciari:2011ma} in experiments are collections of collimated showers of hadrons within a jet cone.
Neglecting nonperturbative effects through hadronization, jets reconstructed through jet shower partons and final hadrons are equivalent and have approximately the same energy. Since interaction in pQCD between an energetic parton and the thermal medium is dominated by small-angle scatterings, a large fraction of thermal recoil  partons from jet-medium interaction fall within the jet cone and therefore should become a part of the reconstructed jet. These thermal recoil  partons within the jet cone can contribute to a non-negligible fraction of the total jet energy and modify both the jet transverse profile (transverse energy distribution inside the jet cone) and fragmentation functions (particle distributions in longitudinal momentum inside the jet cone) \cite{Wang:2013cia}. In addition, a diffusion wake generated behind a propagating parton effectively modifies the underlying fluid background and therefore also affects jet reconstruction, the final jet energy, and the jet profile. These effects can all be studied within the LBT model and are important for understanding the medium modified jet cross section and jet profiles \cite{Chatrchyan:2013kwa}.

In this paper, we describe in detail the LBT model that has been developed over the last few years and improvements to the early versions that have been used for the study of jet-induced medium excitations \cite{Li:2010ts,Wang:2010yz}, jet transport, and medium modifications \cite{Wang:2013cia,Wang:2014hla}. We also employ the model to study elastic energy loss through multiple scattering, propagation of jet-induced medium excitations, and their contributions to parton distribution and jet profiles of reconstructed jets using the anti-$k_T$ algorithm in FASTJET \cite{Cacciari:2011ma}.

Instead of using the small-angle approximation for parton scatterings in the early versions of the LBT model, we have included the complete set of $2\rightarrow 2$ ``elastic" parton scattering processes with full scattering matrix elements including large-angle scatterings, annihilation processes, and flavor conversion. These are necessary for more accurate description of parton transport in medium. Simulations of parton transport according to the Boltzmann equation in the LBT model are based on local rates of scattering in a medium that evolves according to relativistic hydrodynamic equations.  We keep track of medium recoil partons from each parton-medium scattering and neglect interactions among jet shower partons and thermal recoil  partons, wherein lies the meaning of ``linear" in LBT. The model is therefore Lorentz covariant and can avoid difficulties of implementing the test-particle method in a full-fledged parton transport model \cite{Xu:2004mz}. Furthermore, it also allows us to study jet transport and jet-induced medium excitations based on event-by-event simulations with full fluctuations and correlations.

This article is devoted to a general description and test of elastic scatterings in the LBT model and the corresponding jet transport and jet-induced medium excitations in a uniform QGP medium. The final medium modification of jets and the effect of jet-induced medium excitations will depend strongly on the inelastic parton-medium scatterings and the dynamical expansion of the medium. These will be described in the future when we carry out realistic studies of jet transport and medium modification in high-energy heavy-ion collisions.

The remainder of this paper is organized as follows. We first give a detailed description of the LBT model in Sec.~II. We compare numerical and approximate analytic results for local parton scattering rates for different channels, parton energies, and local temperatures. Differential scattering rates as functions of the final-state momentum and scattering angle of the medium recoil parton from Monte Carlo simulations are also tested against results calculated with numerical integrations. In Sec.~III, we study the elastic energy loss from multiple scattering and in particular the energy and time or length dependence in a static and uniform QGP medium. Time evolution of the transverse momentum distribution of the leading parton is also investigated. Propagation of  medium recoil partons in terms of jet-induced medium excitations are investigated in Sec. IV. A supersonic shock wave along with a diffusion wake is shown to arise from the propagation of the energy and momentum deposited by the propagating energetic parton. In Sec.~V, we investigate contributions from  medium recoil partons to the reconstructed jet energy using the anti-$k_T$ algorithm in FASTJET jet finding package and study the influence of  medium  recoil partons to the jet energy profile and parton distributions. Finally, a summary and discussions are given in Sec. VI.

\section{The linear Boltzmann transport model}

Considering only $2 \rightarrow 2$ parton scattering, jet transport and propagation in a QGP medium can be approximately described by a set of linear Boltzmann equations,
\begin{equation}
\label{LBT}
\begin{split}
p_1\cdot\partial f_a(p_1) &= -\int \negthickspace \frac{d^3p_2}{(2\pi)^3 2E_2} \negthickspace \int \negthickspace
\frac{d^3p_3}{(2\pi)^3 2E_3} \int \negthickspace \frac{d^3p_4}{(2\pi)^3 2E_4} \\
\sum_{b(c,d)}&\frac{g_b}{2}\left[ f_a(p_1)f_b(p_2)-f_c(p_3)f_d(p_4)\right] \, |M_{ab\rightarrow cd}|^2  \\
&\times S_2(s, t, u)(2\pi)^4\delta^4(p_1+p_2-p_3-p_4),
\end{split}
\end{equation}
where the sum is over the flavor of the initial thermal parton ($b$) and all possible scattering channels $a+b\rightarrow c+d $
for different flavors of the final partons ($c$,$d$). $|M_{ab\rightarrow cb }|^2$ is the corresponding matrix element \cite{Eichten:1984eu} that is averaged (summed) over the initial (final)  spin and color as a function of standard Mandelstam variables $s$, $t$, and $u$. We have neglected the effect of quantum statistics in this study. The phase-space distributions for thermal partons in the QGP medium $f_i$ $(i=b,d)$ are the Bose-Einstein distribution for gluons and Fermi-Dirac distribution for quarks and antiquarks with a local temperature $T$ and fluid velocity $u=(1, \vec{v})/\sqrt{1-\vec{v}^2}$.  For $N_c=3$ number of colors in QCD, the spin and color degeneracy factors are $g_g=2\cdot(N_c^2-1)=16$ and $g_q=2\cdot N_c = 6$ for gluons and  each flavor of quark or antiquark,  respectively. The propagating parton before and after each scattering is assumed to follow a classical trajectory with $f_i=(2\pi)^3\delta^3(\vec{p}-\vec{p_i})\delta^3(\vec{x}-\vec{x_i}-\vec{p_i}t/E_i)$ $(i=a,c)$ as the phase-space density.

Most matrix elements $|M_{ab\rightarrow cb }|^2$ for two-parton scattering diverge at small angle, $u, t\rightarrow 0$, for massless partons. Such divergences disappear when quasi-particle modes due to hard-thermal-loop (HTL)  corrections are taken into account in the calculation of scattering rates and transport coefficients \cite{Arnold:2001ba,Arnold:2002ja}. While an exact formulation of the Boltzmann transport within HTL pQCD is difficult, we introduce a screening mass for light quarks and gluons in matrix elements for two-parton scattering. This is the same approach as in Refs.~\cite{Zapp:2008gi,Auvinen:2009qm}. This is equivalent to
introducing a Lorentz-invariant regularization condition,
\begin{equation}
\label{cutoff}
S_2(s, t, u) = \theta(s\ge 2\mu_{D}^2)\theta(-s+\mu_{D}^2\le t\le -\mu_{D}^2),
\end{equation}
in the interaction kernel in the Boltzmann transport equations [Eq.~(\ref{LBT})],
where $\mu_{D}$ is the Debye screening mass,
\begin{equation}
\label{cutoff}
\mu_{D}^2 = \frac{g^2 T^2}{3} (N_c + \frac{N_f}{2} ) \ ,
\end{equation}
for $N_f=3$ number of active quark flavors in the QGP.

Matrix elements for parton scattering we use in this paper are leading-order pQCD results. The strong coupling constant
$\alpha_{\rm s}=g^{2}/4\pi$ will be kept at a constant value and should be determined through comparisons to the experimental data.
Since we will only consider parton propagation in a uniform medium in this study, we set $\alpha_{\rm s}=0.3$ throughout the paper.

\subsection{Scattering rates}

\begin{figure}
\includegraphics[width=8.5cm,bb=15 150 585 687]{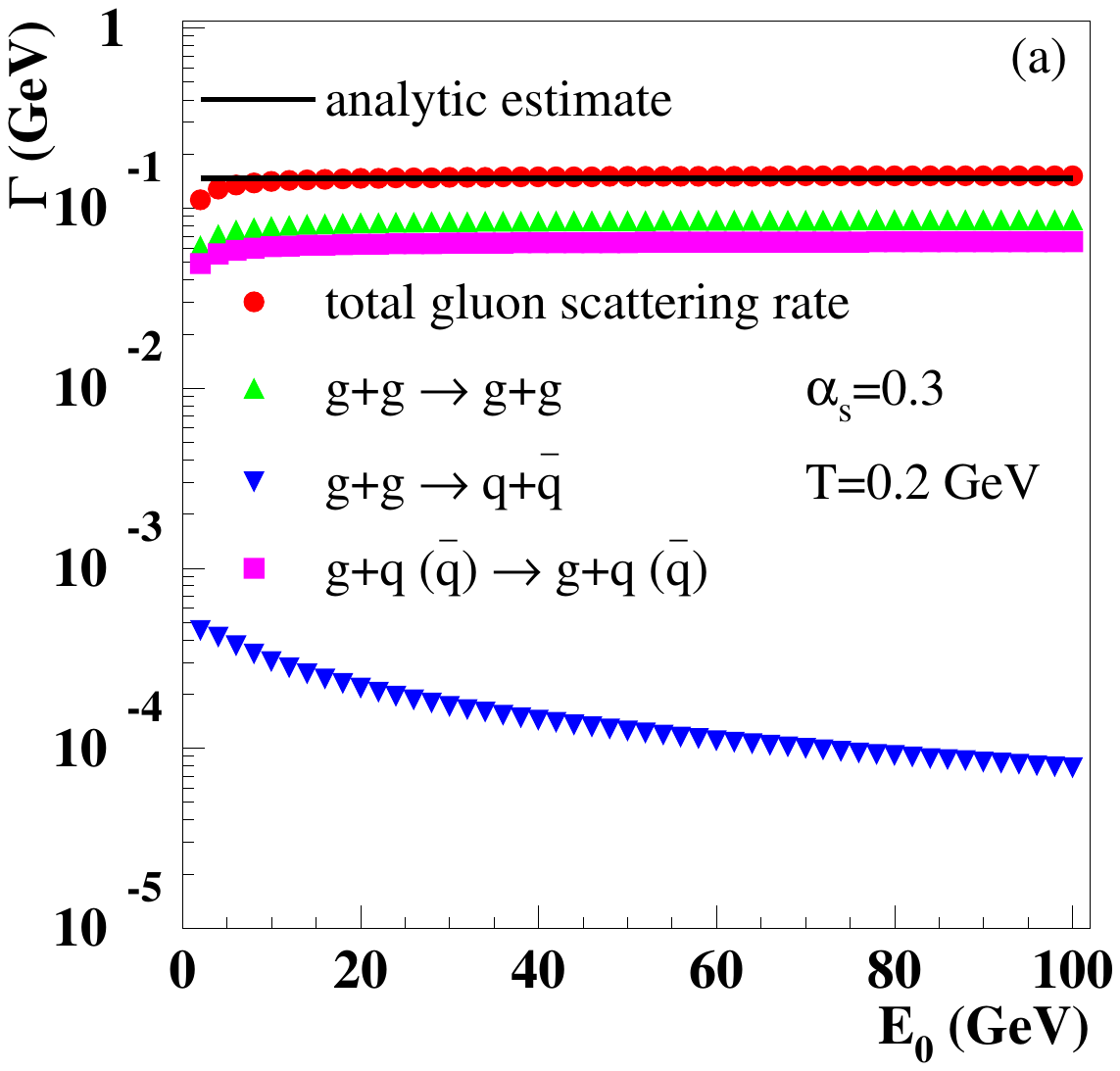}\\
\includegraphics[width=8.5cm,bb=15 150 585 687]{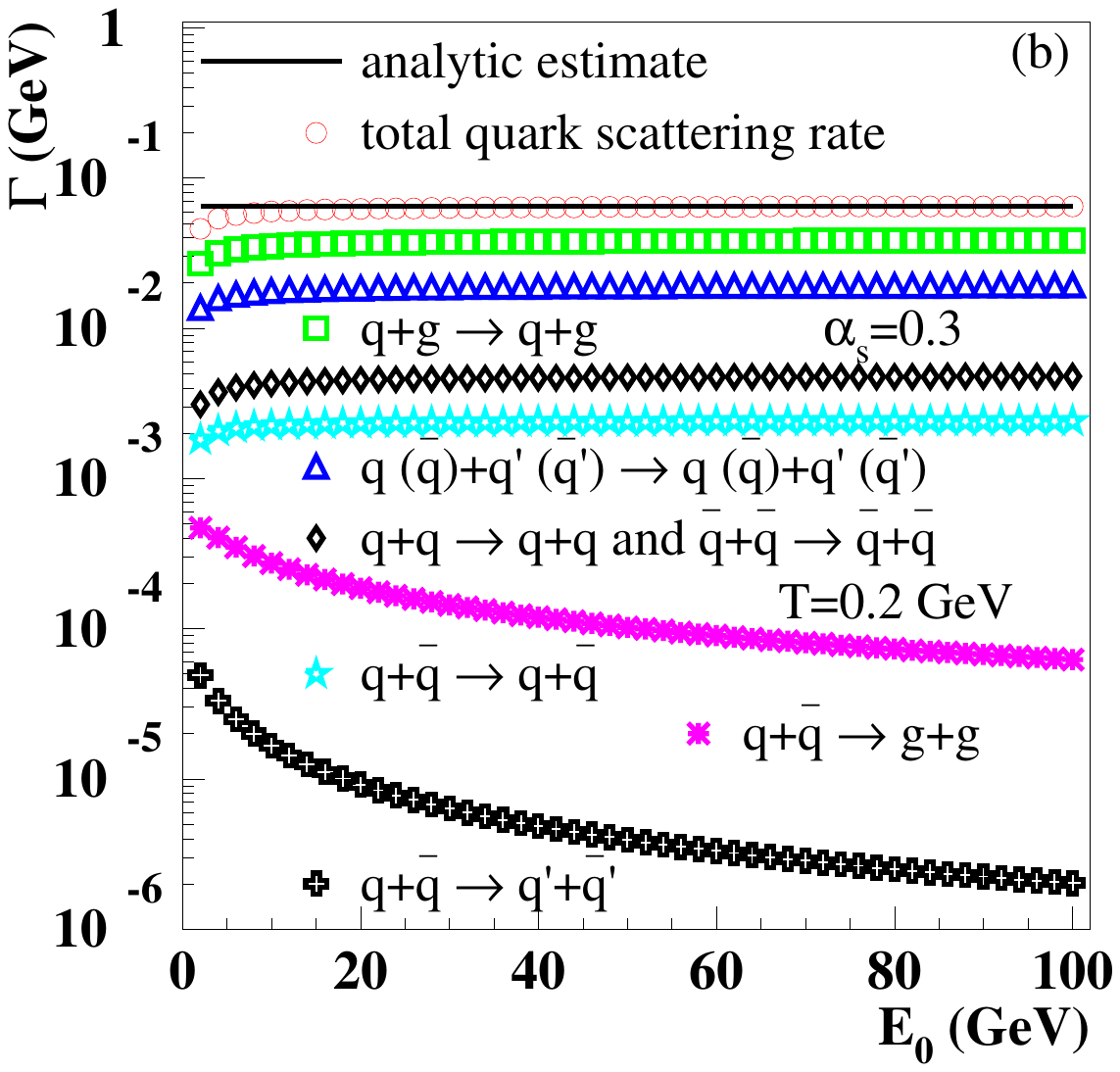}\\
\caption{(Color online) Flavor-summed scattering rates of different processes  \(\Gamma_{ab\rightarrow cd}\) and total scattering rates \(\Gamma_a\) of (a) a gluon or (b) a quark (antiquark) as a function of the parton energy \(E_0\) in a static and uniform QGP medium. Analytic estimates for the total rates discussed in the text are shown as solid lines.}
\label{fig:rates_vs_E}
\end{figure}

For Monte Carlo simulations of the Boltzmann transport equation, we need to evaluate the rate for a hard parton of type $a$ scattering with a thermal parton of type $b$ via a specific channel $a+b\rightarrow c+d $,
\begin{equation}
\label{singlerate}
\begin{split}
\Gamma_{ab\rightarrow cd} = &\frac{g_b}{2E_1} \int \negthickspace \frac{d^3p_2}{(2\pi)^3 2E_2} \negthickspace \int \negthickspace \frac{d^3p_3}{(2\pi)^3 2E_3} \int \negthickspace \frac{d^3p_4}{(2\pi)^3 2E_4} \\
&\times f_b (p_2) \, |M|_{ab\rightarrow cd}^2(s,t,u) \, S_2(s, t, u) \\
&\times (2\pi)^4 \delta^{(4)}(p_1+p_2-p_3-p_4).
\end{split}
\end{equation}
Summing over all types of initial thermal partons ($b$) and all possible channels with different types of final-state partons ($c$,$d$),
we can obtain the total scattering rate for an energetic parton $a$ with thermal partons in a thermal QGP medium:
\begin{equation}
\label{totalrate}
\Gamma_a = \sum_{b,(cd)} \, \Gamma_{ab\rightarrow cd}.
\end{equation}
These scattering rates for different parton types and different channels will be used in Monte Carlo simulations of the LBT model. Shown in Fig.~\ref{fig:rates_vs_E} are the total scattering rates for an energetic gluon (upper panel) and quark (lower panel) as a function of the incident energy $E$ in a thermal QGP medium with zero baryon chemical potential at a constant temperature $T=200$ MeV. Also shown are contributions from different channels. Dominant contributions to the total scattering rate of a gluon come from gluon-gluon and gluon-quark scattering. Contributions from gluon-gluon to quark-antiquark conversion are about two orders of magnitude smaller. The dominant channel to the quark scattering rate is from quark-gluon Compton scattering followed by quark-quark scattering. Contributions from quark-antiquark annihilation are also many orders of magnitude smaller.

As shown in Fig.~\ref{fig:rates_vs_E}, scattering rates for $s$-channel annihilation processes decrease with the incident energy as $\Gamma=\sigma\rho\sim \alpha_{\rm s}^2 T^2/E$. Total parton scattering rates are dominated by $t$-channel gluon-gluon, gluon-quark, and quark-quark scattering processes, whose rates are approximately independent of the incident energy $E$.

\begin{figure}
\includegraphics[width=8.5cm,bb=15 150 585 687]{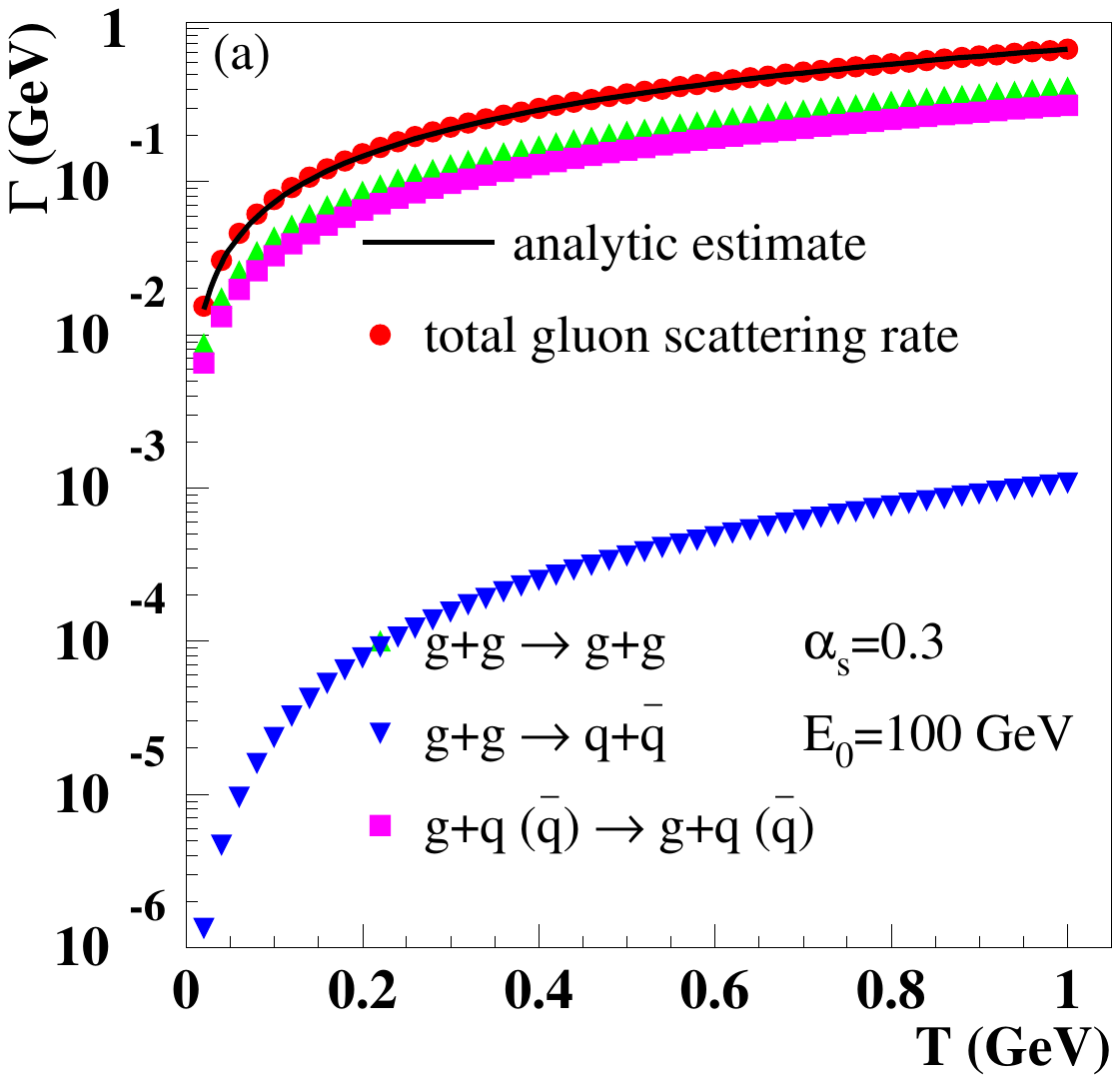}\\
\includegraphics[width=8.5cm,bb=15 150 585 687]{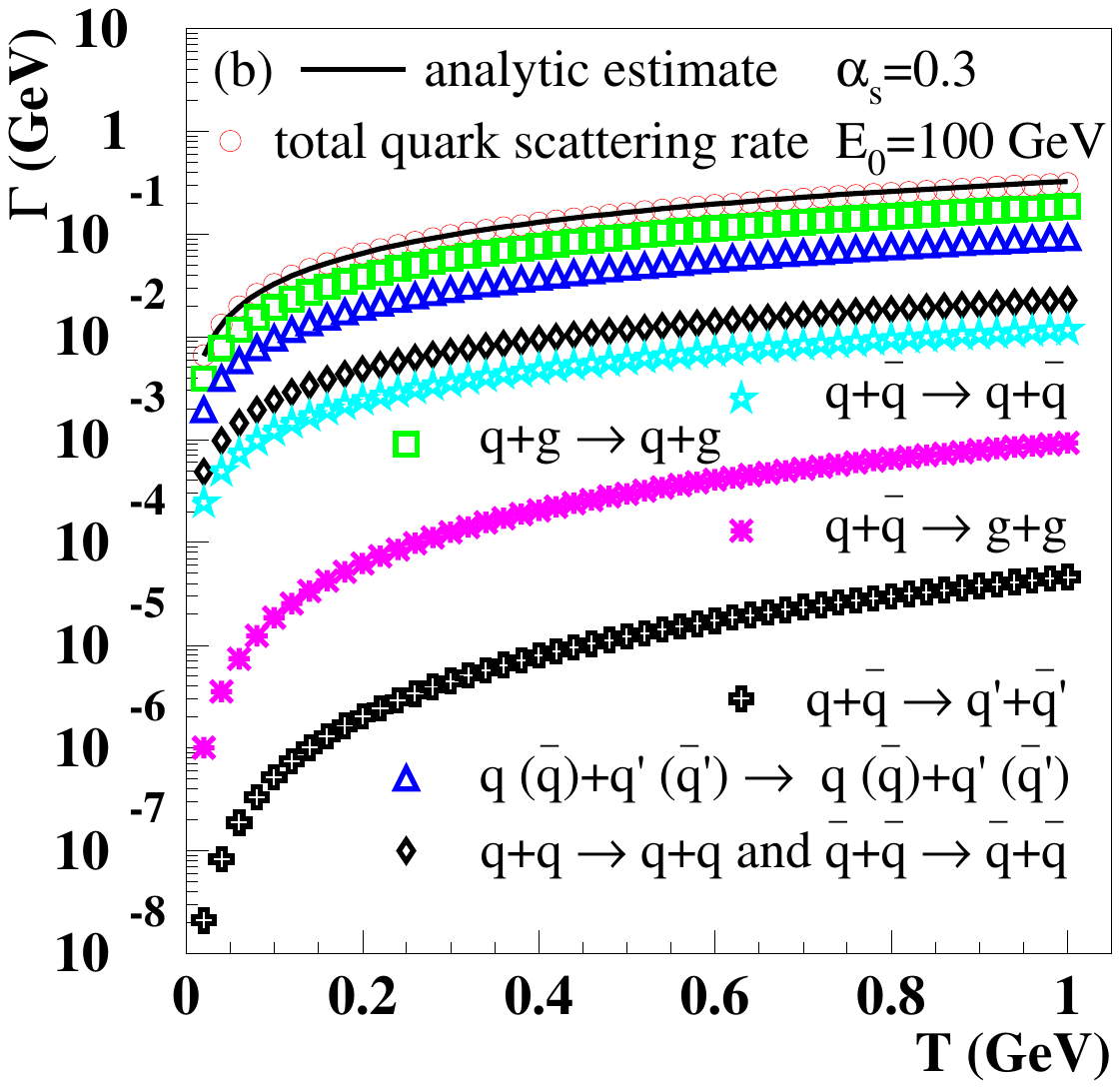}\\
\caption{(Color online) Flavor-summed scattering rates of different processes \(\Gamma_{ab\rightarrow cd}\) and the total scattering rates \(\Gamma_a\) of (a) a  gluon or (b) a quark (antiquark) as a function of the temperature \(T\) in a static and uniform QGP medium. Analytic estimates for the total rates discussed in the text are shown as solid lines.}
\label{fig:rates_vs_T}
\end{figure}

In the high-energy limit $E\gg T$, $t$-channel gluon and quark scattering cross sections can be approximated by their small-angle limits ,
\begin{equation}
\frac{d\sigma_{ab}}{dq_\perp^2} \approx C_{ab} \frac{2\pi\alpha_{\rm s}^2}{q_\perp^4}  \left(C_{gg}=\frac{9}{4}, C_{qg}=1, C_{qq}=\frac{4}{9}\right).
\label{eq-small-el}
\end{equation}
Scattering rates for energetic partons in a thermal QGP medium are then \cite{Auvinen:2009qm,Wang:1996yf},
\begin{eqnarray}
\label{analytic_gammag}
\Gamma_g &\approx& \sum_{b=g,q_i,\bar q_i} \Gamma_{g b  \rightarrow g b}
\approx 42C_A\zeta(3) \frac{\alpha_s^2T^3}{\pi \mu_D^2},\\
\label{analytic_gammaq}
\Gamma_{q} &\approx& \sum_{b=g,q_i,\bar q_i}\Gamma_{q b  \rightarrow q b }
 \approx 42C_F\zeta(3) \frac{\alpha_{\rm s}^2T^3}{\pi \mu_D^2},
\end{eqnarray}
where $\zeta(3)\approx 1.202$ is the Ap\'{e}ry's constant, $C_A=N_c$ and $C_F=(N_c^2-1)/2N_c$. With the Debye screening mass given in Eq.~(\ref{cutoff}), the above scattering rates are independent of the incident energy and proportional to the temperature $T$.  As shown in Fig.~\ref{fig:rates_vs_E} as solid lines, the above analytic estimates of total parton scattering rates agree with numerical results very well for $E\gg T$. We also show in Fig.~\ref{fig:rates_vs_T} both numerical results (symbols) and analytic estimates (solid lines) of total scattering rates and contributions from different channels as functions of the temperature in a thermal QGP medium for a fixed parton incident energy $E=100$ GeV. As given in the above analytic estimates, total rates as well as rates for $t$ channels increase linearly with the temperature $T$, while rates for $s$ channels increase quadratically with $T$. Note that the parton scattering rates given here are only valid in a medium above the QCD phase transition temperature $T>T_c$. For a rapid crossover as indicated by recent lattice QCD simulations \cite{Bazavov:2014pvz},  the effective degrees of freedom around the phase transition temperature $T_c\approx 155 $ MeV become ambiguous. One has to rely on model assumptions to  treat  parton medium interaction \cite{Wang:2013cia,Chen:2010te,Chen:2011vt}.

\subsection{Differential rates}

To sample the initial thermal and final parton four-momenta in a $2 \rightarrow 2$ scattering, we choose the $z$ axis as the direction of the incident parton $a$ with its four-momentum $p_1=(E_1,\vec 0_\perp, p_{1z})$. One can integrate out the four-dimensional $\delta$ function in Eq.~(\ref{singlerate}) and express the the scattering rate for a given channel in a differential form:
\begin{equation}
\label{scattrate2}
\begin{split}
\Gamma _{ab\rightarrow cd}=&\frac{g_b}{16E_1(2\pi )^4}\int \negthickspace d\theta _2\int \negthickspace d\theta_3\int \negthickspace d\phi _{23}\int \negthickspace dE_3 \\
&\times f_b(E_2,T)\left| M \right|_{ab\to cd}^{2}(s,t,u){{S}_{2}}(s,t,u) \\
&\times \frac{E_2 E_3\sin \theta_2\sin\theta_3}{E_1(1-\cos\theta_{12})-E_3(1-\cos\theta_{23})}, \\
\end{split}
\end{equation}
where
\begin{equation*}
{{E}_{2}}=\frac{{{E}_{1}}{{E}_{3}}(1-\cos {{\theta }_{3}})}{{{E}_{1}}(1-\cos {{\theta }_{12}})-{{E}_{3}}(1-\cos {{\theta }_{23}})},
\end{equation*}
 $\phi_i$ is the azimuth angle,  $\theta_i$ the  polar angle of a parton's momentum $p_i$, and $\phi_{ij}$ and  $\theta_{ij}$  are  the azimuth and polar angles between two partons' momenta $p_i$ and $p_j$, respectively.

\begin{figure}
\includegraphics[width=8.0cm,bb=15 150 585 687]{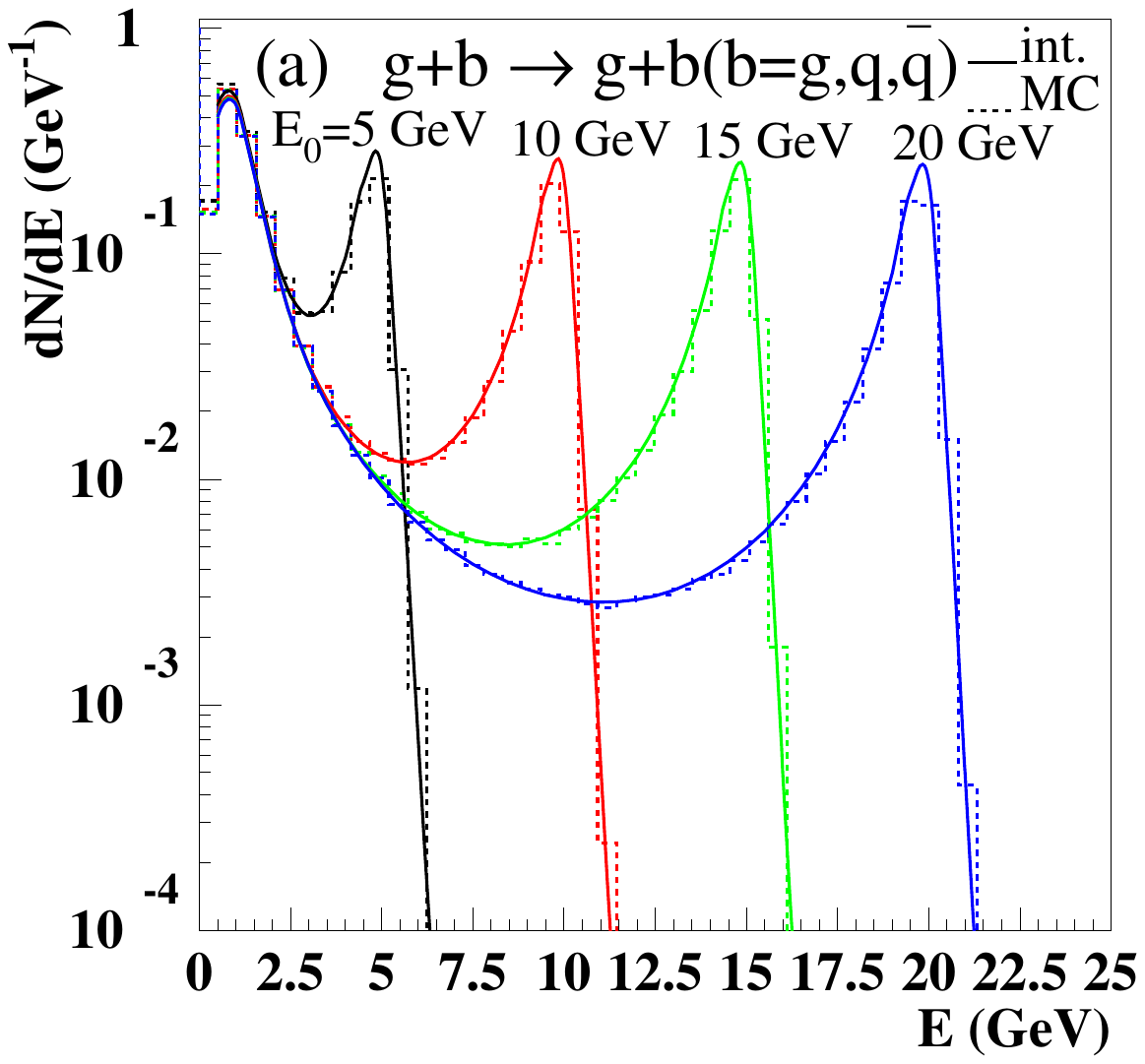}
\includegraphics[width=8.0cm,bb=15 150 585 687]{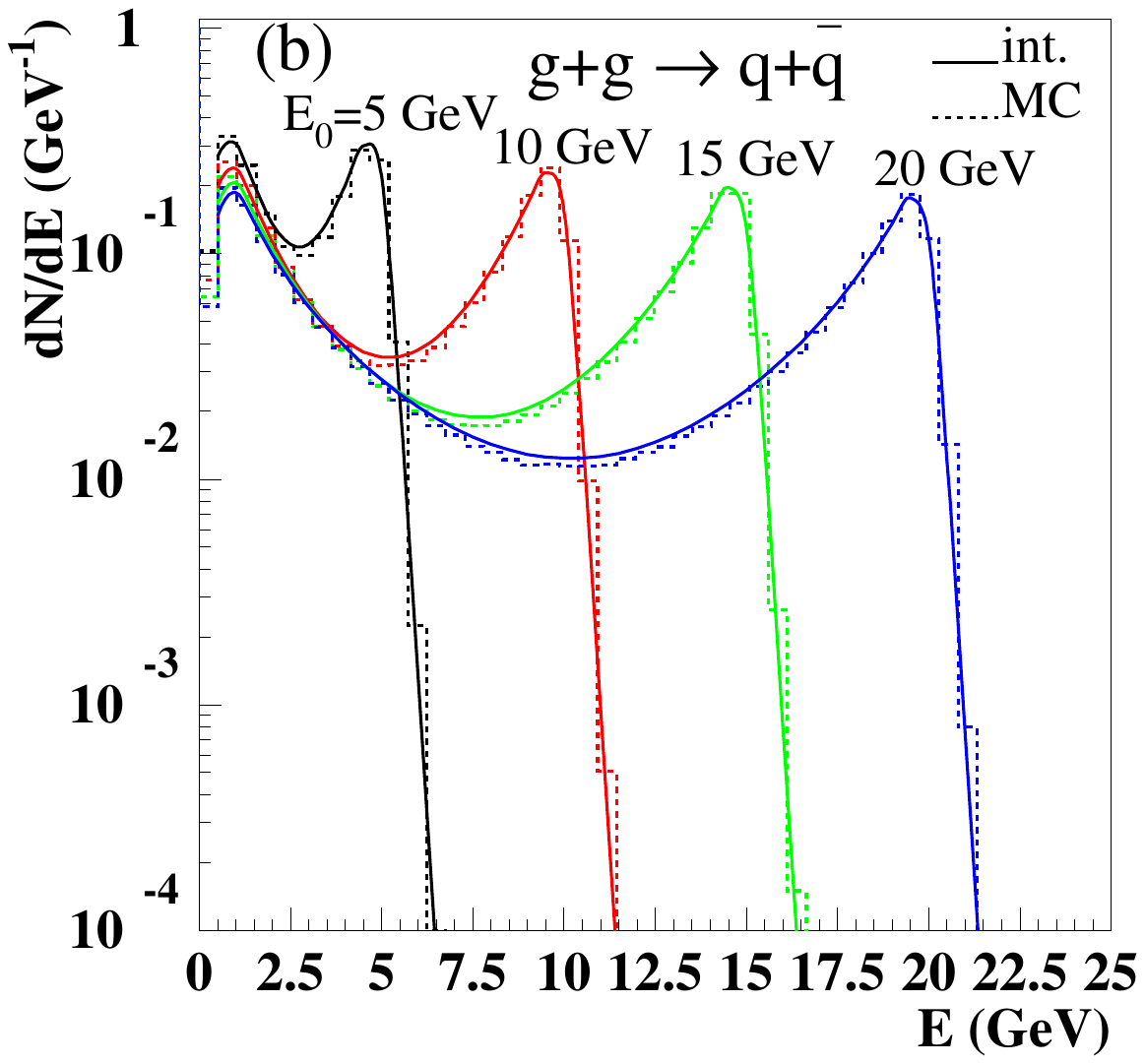}
\caption{(Color online) Energy distributions of the outgoing parton in (a)  $g+ b\rightarrow g + b (b=g,q,\bar q) $  or  (b) $g+g\rightarrow q + \bar q $ scattering processes with different initial parton energy $E_0$ in a thermal QGP medium at a temperature $T=200$ MeV.  Solid lines are results from Eq.~(\ref{scattrate2}) with direct integration while histograms are from the corresponding Monte Carlo simulations.}
\label{fig-e-gluon}
\end{figure}

\begin{figure}
\includegraphics[width=8.0cm,bb=15 150 585 687]{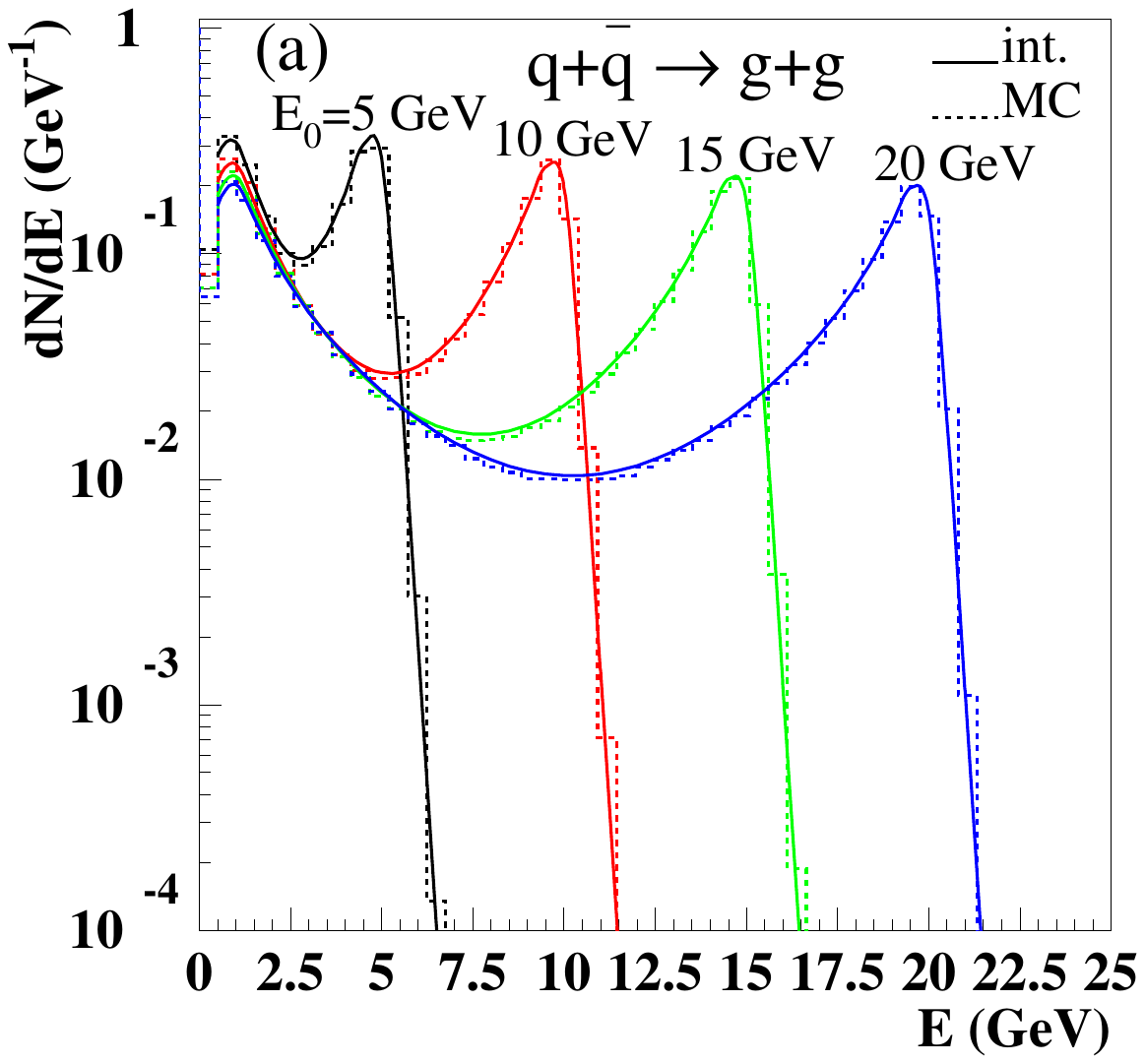}
\includegraphics[width=8.0cm,bb=15 150 585 687]{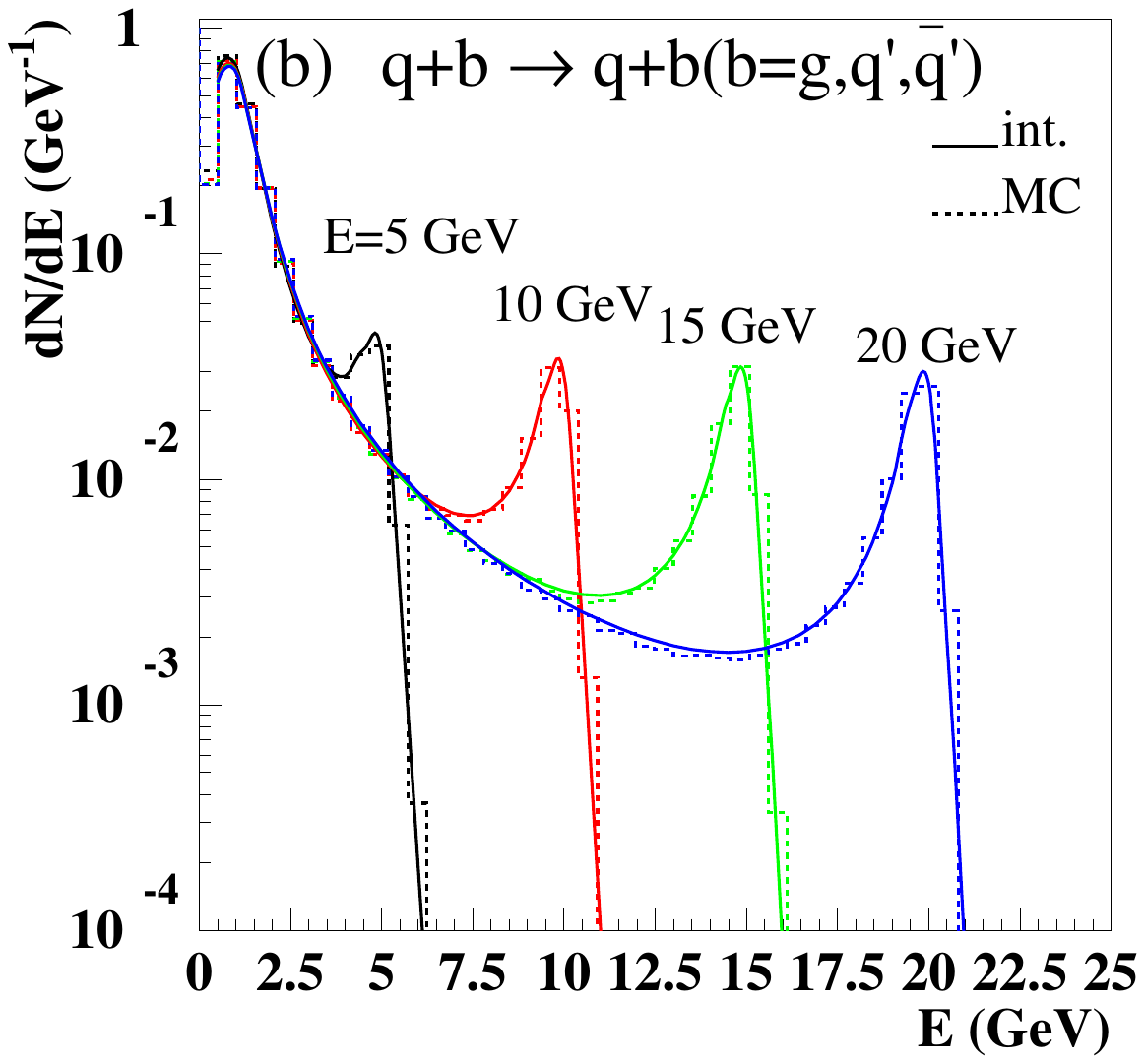}
\caption{(Color online) The same as Fig.~\ref{fig-e-gluon} in (a)  $q+\bar q \rightarrow g+g$ or (b) $q+b \rightarrow q +b (b=g,q\prime,\bar q\prime)$ scattering processes.}
\label{fig-e-quark}
\end{figure}

One can further carry out the integration over the direction of the initial thermal parton $b$'s momentum $(\theta_2,  \phi_{23})$ and obtain the final parton $c$'s angular ($\theta_3$) and energy ($E_3$) distributions. Shown in Figs.~\ref{fig-e-gluon}  and \ref{fig-e-quark} as solid lines are the energy distributions of an outgoing parton in some selected channels of parton-medium interaction in a thermal QGP with a temperature $T=200$ MeV.  The gluon-medium interaction is dominated by $t$ and $u$-channel gluon-gluon and gluon-quark(antiquark) scattering. The outgoing gluon can either be the deflected incoming gluon which carries most of the incident energy $E_0$  minus elastic energy loss $\sim \mu_D^2/T$ or the thermal recoil  gluon whose average energy is its initial thermal energy $\sim T$ plus the elastic energy transfer from the scattering. The energy spectrum of the outgoing gluon therefore has two peaks at $E\sim T$ and $E_0$ as shown in the upper panel of Fig.~\ref{fig-e-gluon}.  Large-angle and $s$-channel scatterings contribute to the valley region between these two peaks which is suppressed by a factor of $\mu_D^2/E_0T$. The energy spectrum of the outgoing gluon does not vanish at the incident energy $E_0$. It instead has an exponential fall-off with a slope as given by the temperature, indicating that the energetic incident parton can also gain energy due to its interaction with thermal partons.

The energy spectra of the outgoing quark in  the quark-antiquark pair production $g+g\rightarrow q + \bar q$ (lower panel of Fig.~\ref{fig-e-gluon})  and the outgoing gluon in the quark-antiquark annihilation $q+\bar q\rightarrow g+g$ process (upper panel of Fig.~\ref{fig-e-quark})  have a similar structure with two peaks at $E\sim T$ and $E_0$. This structure, however, arises mainly from the $u$ and $t$ channels of the pair production or annihilation process in which the two outgoing partons have an equal probability to carry most of the incident energy. Contributions from the $s$ channels in these processes are suppressed by a factor of $\mu_D^2/TE_0$ with a flat energy distribution between $E\sim T - E_0$.

To illustrate the typical energy spectra of thermal recoil   partons from parton-medium interaction we show in the lower panel of Fig.~\ref{fig-e-quark} the energy distribution of the recoil parton $b$ in $q+b \rightarrow q+b$ ($b=g,q^\prime,\bar q^\prime$) processes. For an energetic quark $q$, these processes are dominated by $t$ channels in which recoil partons carry an energy $E_b\sim q_T^2/T$. The recoil parton spectra should have a power law behavior $dN/dE_b \sim 1/q_T^4 \sim 1/E_b^2$ with a peak at $E_b\sim T+ \mu_D^2/T$.  The recoil gluon spectra in the $s$ and $u$ channels in the quark-gluon Compton scattering $q+g\rightarrow q+g$, however, still has a peak at $E_b\sim E_0-\mu_D^2/T$, but is suppressed by a factor of $\mu_D^2/E_0T$ relative to the $t$ channel.

\begin{figure}
\includegraphics[width=8.0cm,bb=15 150 585 687]{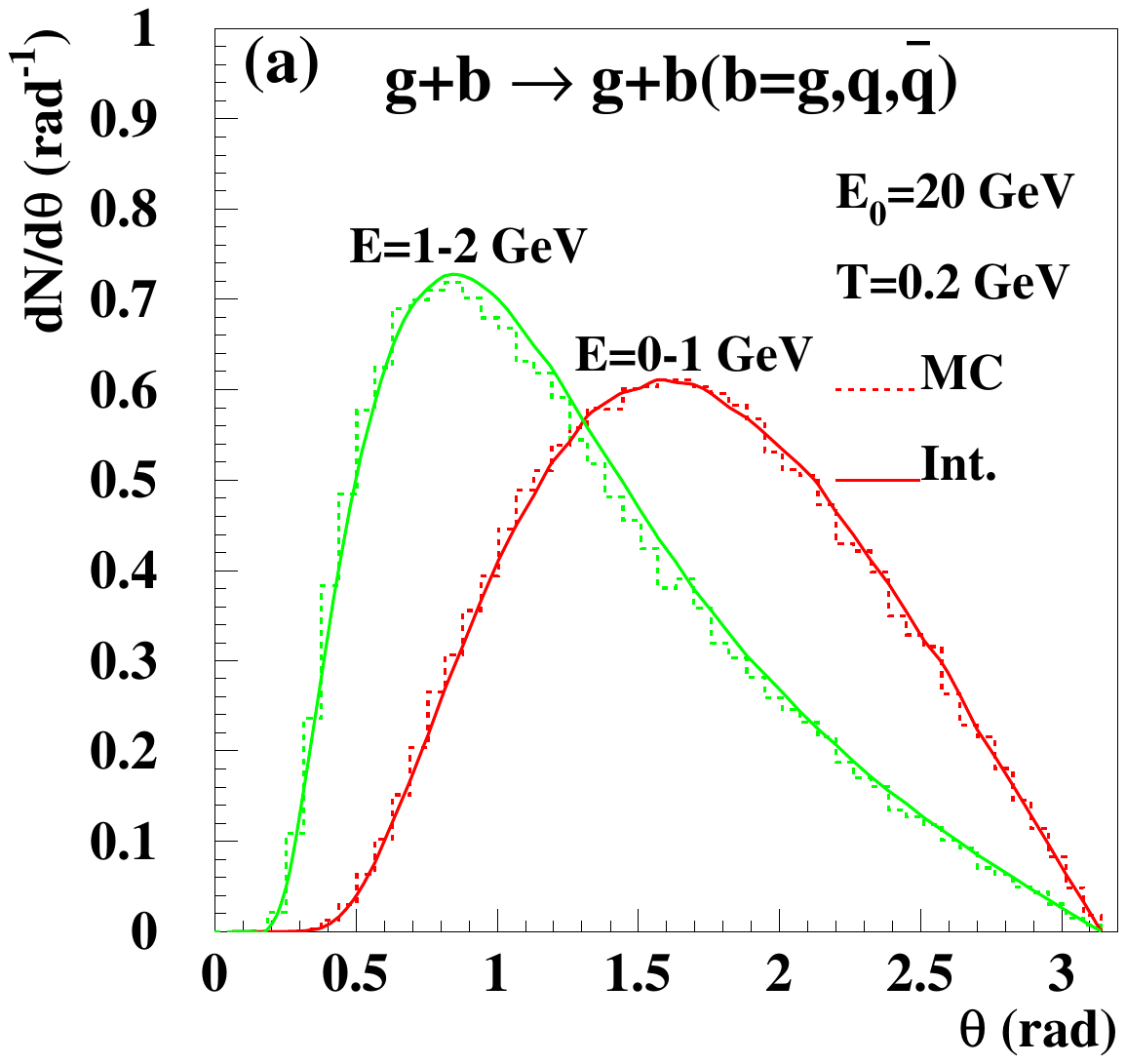}
\includegraphics[width=8.0cm,bb=15 150 585 687]{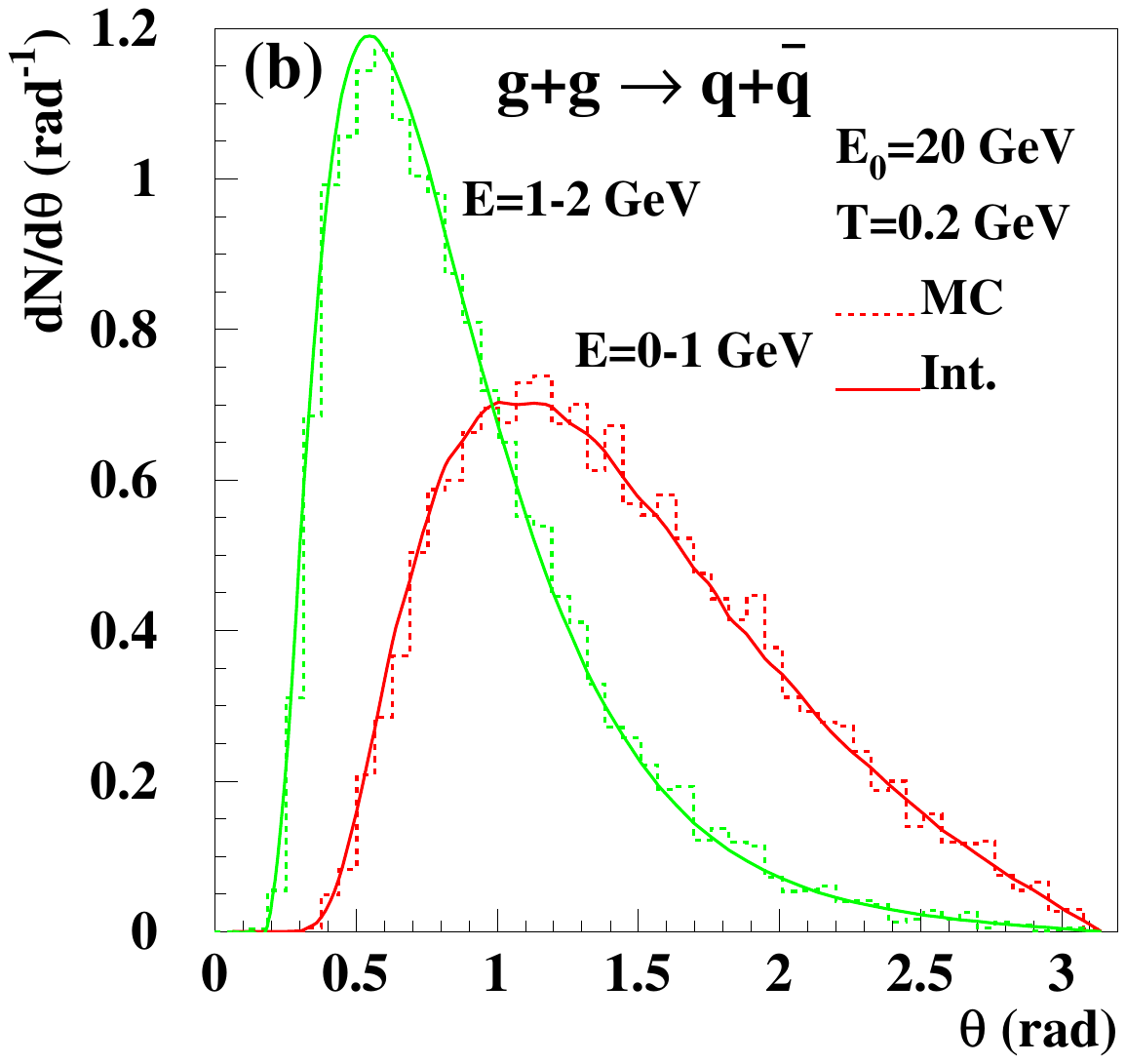}
\caption{(Color online) Polar angle distributions of the thermal recoil  parton of different energies relative to the incoming parton direction after (a) $g+ b\rightarrow g + b \, (b=g,q,\bar q) $ or (b) $g+g\rightarrow q + \bar q $ scattering processes with initial parton energy $E_0=20$ GeV in a uniform medium at a temperature $T=200$ MeV. Solid lines are for results from the numerical integration and  histograms for LBT Monte Carlo simulations. }
\label{fig-theta-gluon}
\end{figure}

\begin{figure}
\includegraphics[width=8.0cm,bb=15 150 585 687]{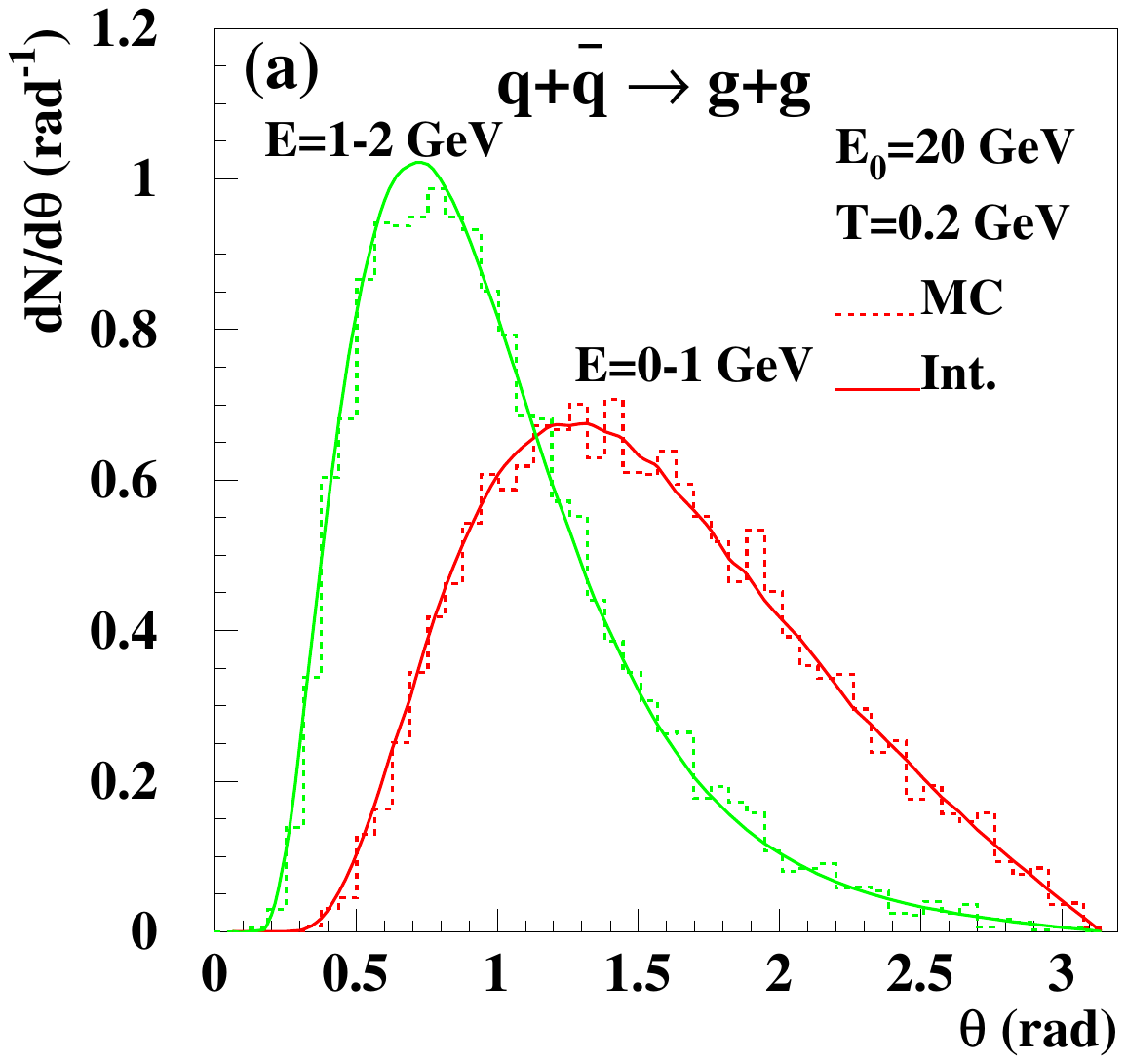}
\includegraphics[width=8.0cm,bb=15 150 585 687]{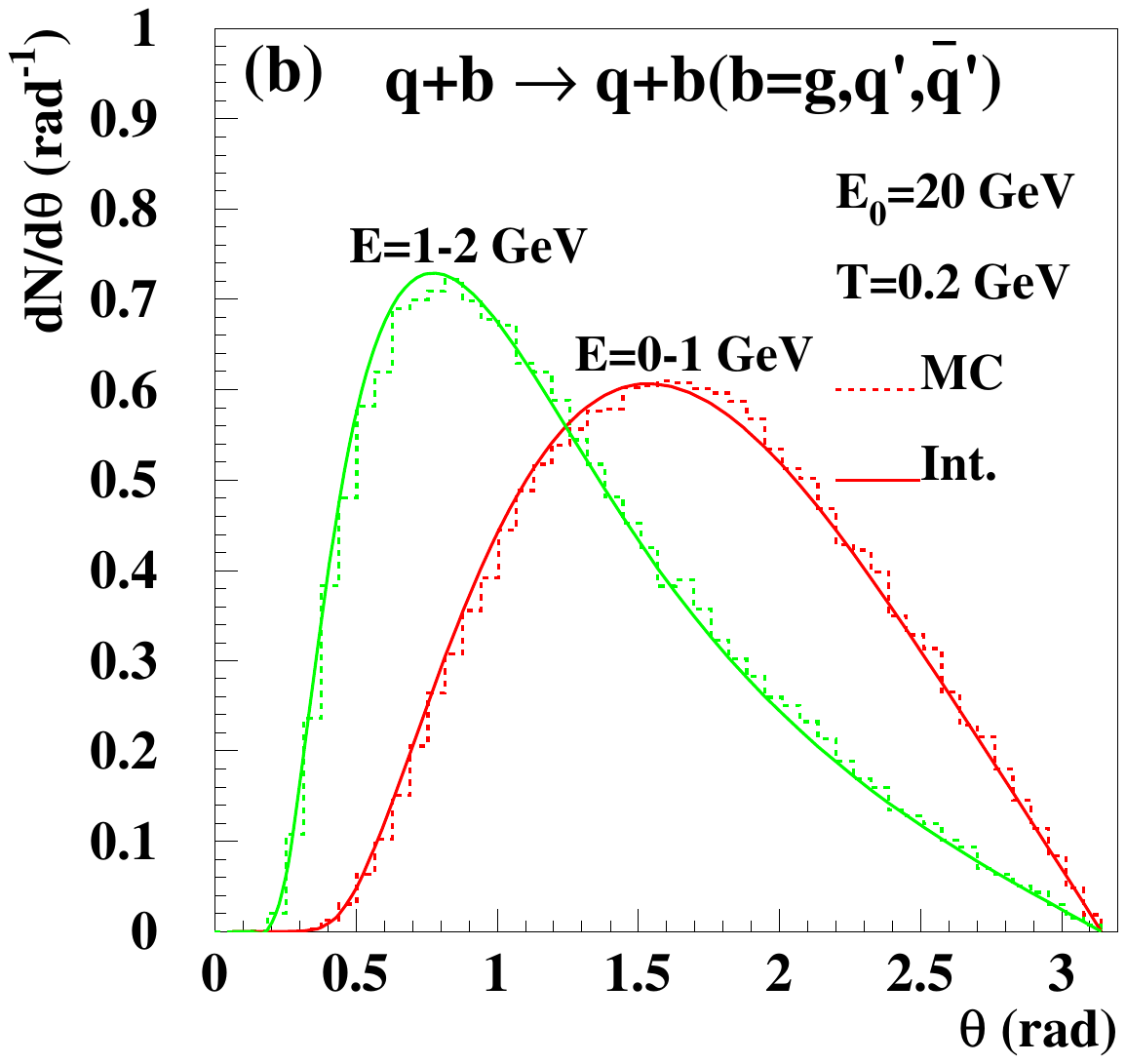}
\caption{(Color online) The same as Fig.~\ref{fig-theta-gluon}  for (a) $ q + \bar q  \rightarrow g+g $  or (b) $q+ b\rightarrow q + b \, (b=g,q^\prime,\bar q^\prime) $ scattering processes.}
\label{fig-theta-quark}
\end{figure}

Because of the dominance of $t$ and $u$ channels of the small-angle scattering, only one of the outgoing partons in parton-medium scattering carries most of the incident energy  for most of the time, whose flavor can be, however, different from the incident parton. Therefore, we will refer to the more energetic parton in the final state of each parton-medium scattering as the leading parton and the other less energetic one as the thermal recoil  parton.

For Monte Carlo simulations in the LBT model, we change the variable of the energy integration in Eq.~(\ref{scattrate2}) from the outgoing parton $c$'s energy $E_3$ to the initial parton $b$'s thermal energy $E_2$. The integrand then can be used to sample the thermal parton $b$'s energy $E_2$, polar angle $\theta_2$ and the outgoing parton $c$'s direction ($\theta_3,\phi_{23}$) with the rejection method. The energy of the outgoing parton $c$ $E_3$ and parton $d$'s four-momentum $p_4$ can be determined by energy-momentum conservation. Shown as histograms in Figs.~\ref{fig-e-gluon}  and \ref{fig-e-quark} are the energy distributions of the outgoing parton in some elected channels of parton-medium scattering. The agreement between Monte Carlo results and those from direct integration in Eq.~(\ref{scattrate2})  provides numerical verifications of the Monte Carlo code for parton-medium scatterings in the LBT model.

For the purpose of discussing jet-induced medium excitation later, we show in Figs.~\ref{fig-theta-gluon} and \ref{fig-theta-quark} distributions in the polar angle of the thermal recoil  parton from parton-medium scatterings via the same selected channels as in Figs.~\ref{fig-e-gluon} and \ref{fig-e-quark} from both Monte Carlo simulations (histograms) and numerical integration of Eq.~(\ref{scattrate2}) (solid lines). The polar angle of the thermal recoil  parton in two different energy ranges, $E=$0-1 and 1-2 GeV, is calculated with respect to the direction of the incident parton with $E_0=20$ GeV. As shown in the figures, thermal recoil  partons with energy significantly larger than the typical thermal energy gain their energy and momentum through elastic scattering and therefore are dragged along in the same direction of the energetic incident parton. Soft thermal recoil  partons with $E$=0-1 GeV, on the other hand, have large angles, almost perpendicular to the incident parton on the average. These features of parton-medium scattering will determine the jet-induced medium excitation due to multiple parton scattering as we will describe later.

\subsection{Multiple scatterings}

In the LBT model, we initiate the Monte Carlo program by calculating parton scattering rates according to Eqs.~(\ref{singlerate}) and (\ref{totalrate}) for a range of incident energies and medium temperatures which are stored in two-dimensional tables for later use. We then approximate the propagation of partons in medium with a discretized time internal $\Delta t$ in the frame of hydrodynamic evolution of the QGP medium. We sample the probability for $n$ number of parton-medium scatterings within a time interval $\Delta t$ according to a Poisson distribution,
\begin{eqnarray}
\label{prob}
P_{a,n}(E,\Delta t)&=&\frac{[\Delta N_a(E)]^n}{n!} e^{-\Delta N_a(E,\Delta t)}, \\
\Delta N_a(E,\Delta t)&=&\Delta t \frac{p\cdot u}{E} \Gamma_a (p\cdot u,T),
\end{eqnarray}
where the scattering rate $\Gamma_a (p\cdot u,T)$ for a parton $a$ is evaluated at the local fluid comoving frame with incident energy $E_u=p\cdot u$ and $u$ is the local fluid four-velocity. After determining the number of parton scatterings $n$, one can assume these $n$ number of scatterings occur sequentially within the time interval $\Delta t$ and energy and momentum are strictly conserved in each scattering  along the classical trajectory.

If we choose the value of time interval $\Delta t$ to be much smaller than the mean-free-path length $\Delta t \ll 1/\max(\Gamma_g,\Gamma_q)$, one can approximate the probability for at least one parton-medium scattering,
\begin{equation}
P_a(E,\Delta t)=1-e^{-\Delta N_a(E,\Delta t)},
\end{equation}
as the probability for one parton-medium scattering during the time interval $\Delta t$.

For each parton-medium scattering, we use fractional rates $\Gamma_{ab\rightarrow cd}/\Gamma_a$ to decide the channel and flavors of the initial thermal parton $b$ and outgoing partons $c$ and $d$. The kinematics of the parton-parton scattering $a+b\rightarrow c+d$ is sampled according to the differential rate in Eq.~(\ref{scattrate2}). The more energetic parton of the two outgoing partons is chosen as the leading parton for the next parton-medium scattering. In the LBT model, all outgoing partons, both the leading and thermal recoil  partons,  in each parton-medium scattering are recorded and are allowed to go through further parton-medium scattering in the subsequent time intervals. Each parton-medium scattering could in principle accompanied by induced gluon bremsstrahlung. We focus mainly on elastic scattering processes in the LBT model in this paper and leave the discussion on the implementation of induced gluon bremsstrahlung for the subsequent publication.

To take into account of the back-reaction in the Boltzmann transport equation (\ref{LBT}), we also record the initial thermal parton $b$ and its four-momentum $p_2$ in each scattering process, which we denote as ``negative'' partons and are allowed to transport further according to the Boltzmann equation. These ``negative'' partons should be subtracted from the final parton spectra and energy-momentum density of the jet-induced medium excitation. Thermal recoil  partons and the``negative" partons are collectively called {\it jet-induced medium partons} in this study.

\section{Transverse momentum broadening and elastic energy loss}

In the LBT model, all channels of $2\rightarrow 2$ scattering are considered for parton-medium interaction in which the flavor of the most energetic parton in the final state can be different from the incident parton. In the LBT Monte Carlo simulations, we designate the more energetic one of the two final partons as the leading parton. One can then follow the propagation of the leading parton and study its  transverse momentum broadening and elastic energy loss.

\subsection{$p_T$ broadening}

During the propagation of an energetic parton in a QGP medium, each parton-medium scattering contributes to the transverse momentum (perpendicular to the initial direction of the propagating parton) of the final leading parton according to matrix elements of the $2\rightarrow 2$ parton-parton scattering, leading to the increase of the averaged transverse momentum squared over time or the transverse momentum broadening. The averaged transverse momentum squared per unit length is often defined as the jet transport parameter $\hat q$ \cite{Gyulassy:1993hr}-\cite{Arnold:2002ja} which characterizes both the local gluon number density and the strength of jet-medium interaction.

\begin{figure}
\includegraphics[width=8.5cm,bb=15 150 585 687]{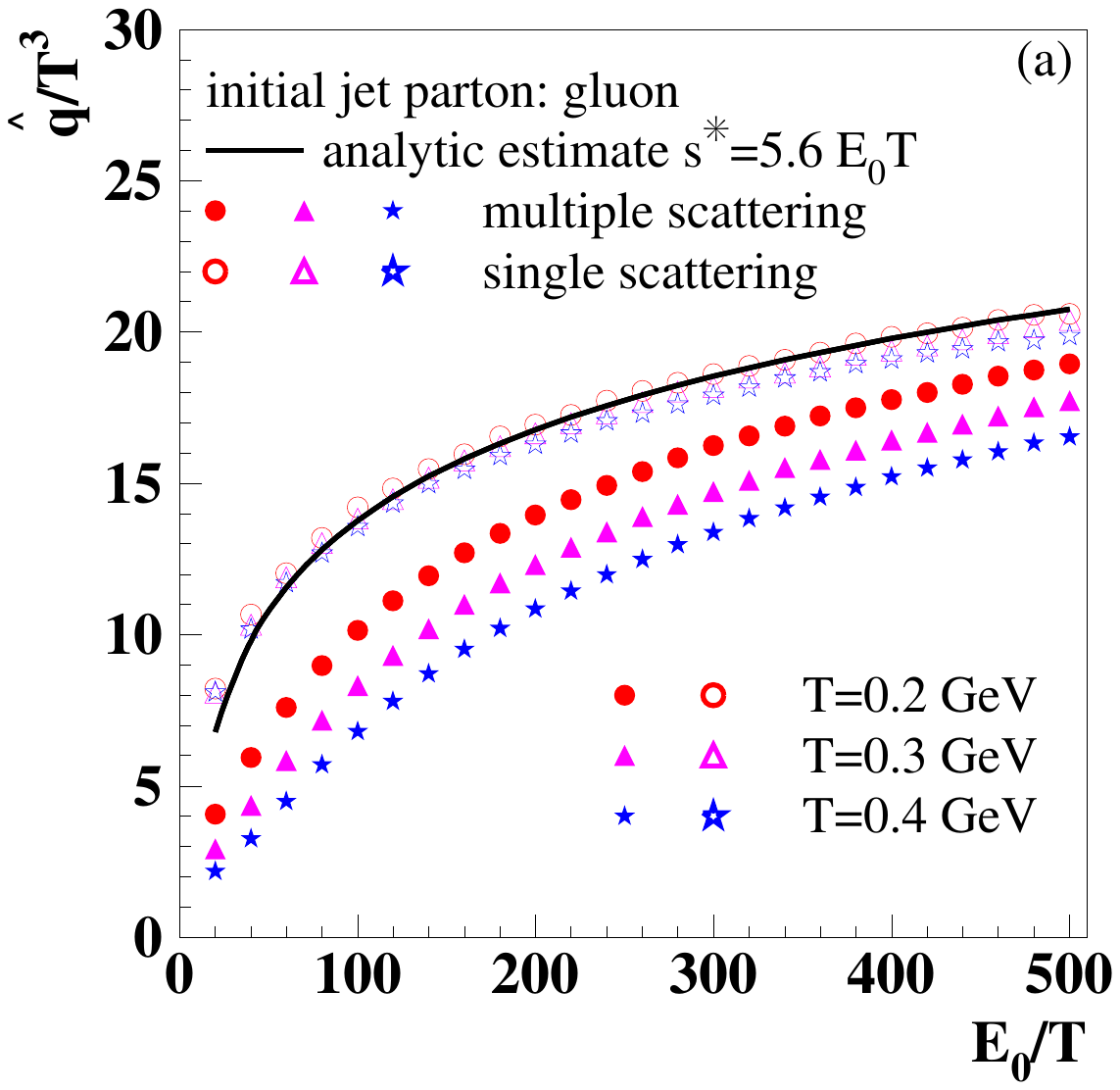}\\
\includegraphics[width=8.5cm,bb=15 150 585 687]{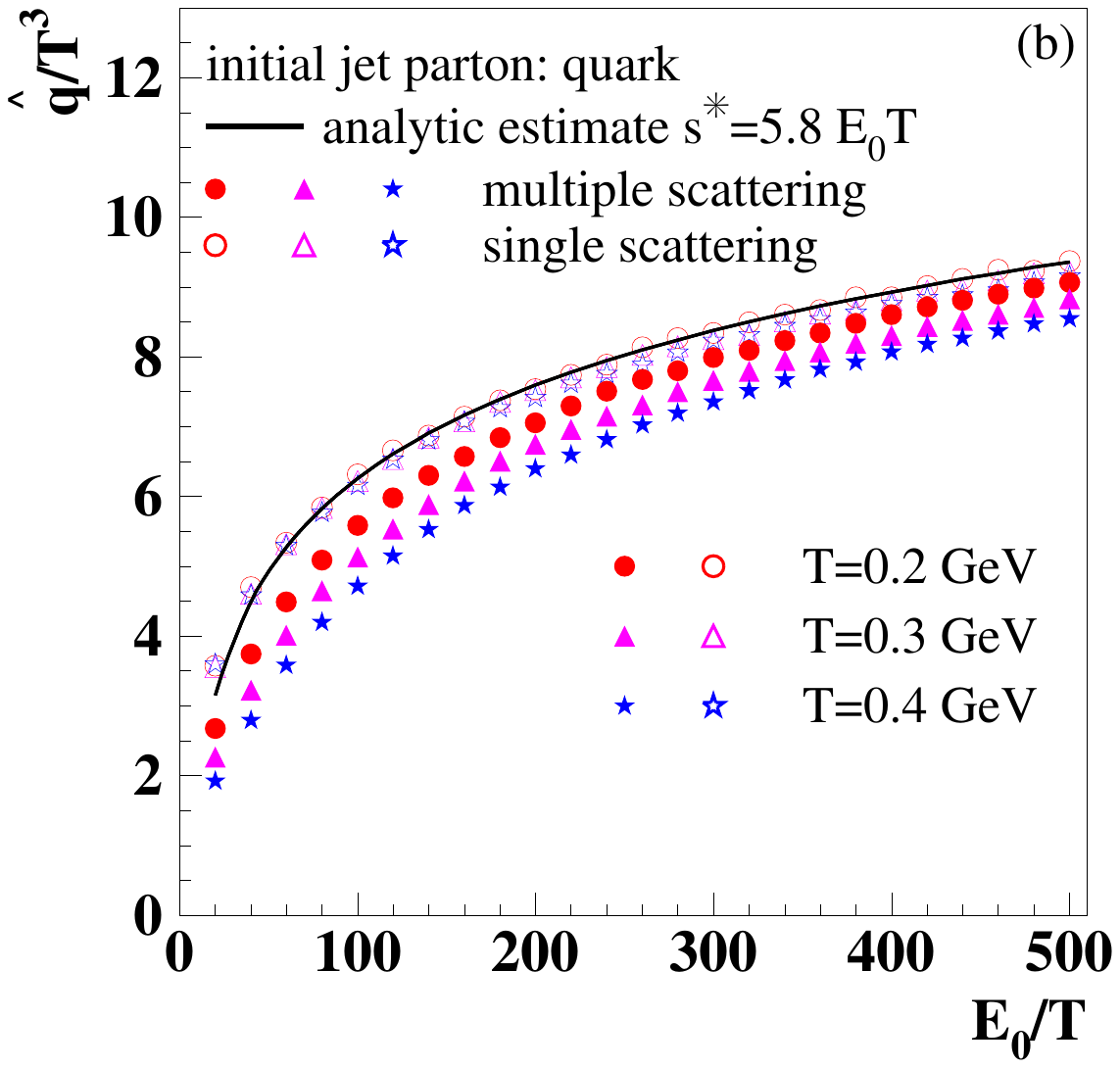}\\
\caption{(Color online) The jet transport parameter $\hat q$ or transverse momentum broadening squared per unit length of the leading parton for (a) an initial gluon or (b) quark with different initial energies going through a single (open symbols) or multiple scatterings (solid symbols)  in a uniform and static QGP medium at different temperatures.  Solid lines are analytic results from a single scattering with the small-angle approximation. }
\label{qhat}
\end{figure}

Employing the small-angle approximation of $2\rightarrow 2$ cross sections in Eq.~(\ref{eq-small-el}), one can estimate the transverse momentum broadening squared per mean-free-path length as,
\begin{eqnarray}
\label{<qperp>}
\hat q_a&=&\langle q_{\bot }^{2}/\lambda\rangle_a =\sum_{b,(cd)} \int_{\mu _{D}^{2}}^{s/4}{dq_{\bot }^{2} \, \frac{d\sigma_{ab\rightarrow cd}}{dq_{\bot }^2}}\rho_b \, q_{\bot }^{2}\nonumber \\
&=&\Gamma_a \langle q_{\bot }^{2}\rangle
\approx C_a \frac{42 \zeta(3)}{\pi} \alpha_{\rm s}^2T^3 \ln (\frac{s^*}{4\mu _D^2}),
\end{eqnarray}
 where $\langle q_{\bot }^{2}\rangle$ is the transverse momentum transfer per scattering which is independent of the parton's flavor, $C_a=C_F=4/3$ for a quark and $C_a=C_A=3$ for a gluon. The variable in the logarithm is defined as $\ln s^*=\langle \ln s\rangle$ averaged over the kinematics of a single parton-medium scattering and $s$ is the center-of-mass energy squared for each parton-parton scattering. One can assume, however, $s^*=2cE_0T$ with the constant $c$ determined from the numerical calculations.

During multiple scatterings in the LBT model, one can follow the propagation of the leading parton and study its final transverse momentum distribution and accumulated transverse momentum broadening.  Shown in Fig.~\ref{qhat} are the transverse momentum broadening per unit length from single and multiple scattering of an energetic parton with different initial energy $E_0$ in a static and uniform QGP at different temperatures over a length of $L=8$ fm.

\begin{figure}
\includegraphics[width=8.5cm,bb=15 150 585 687]{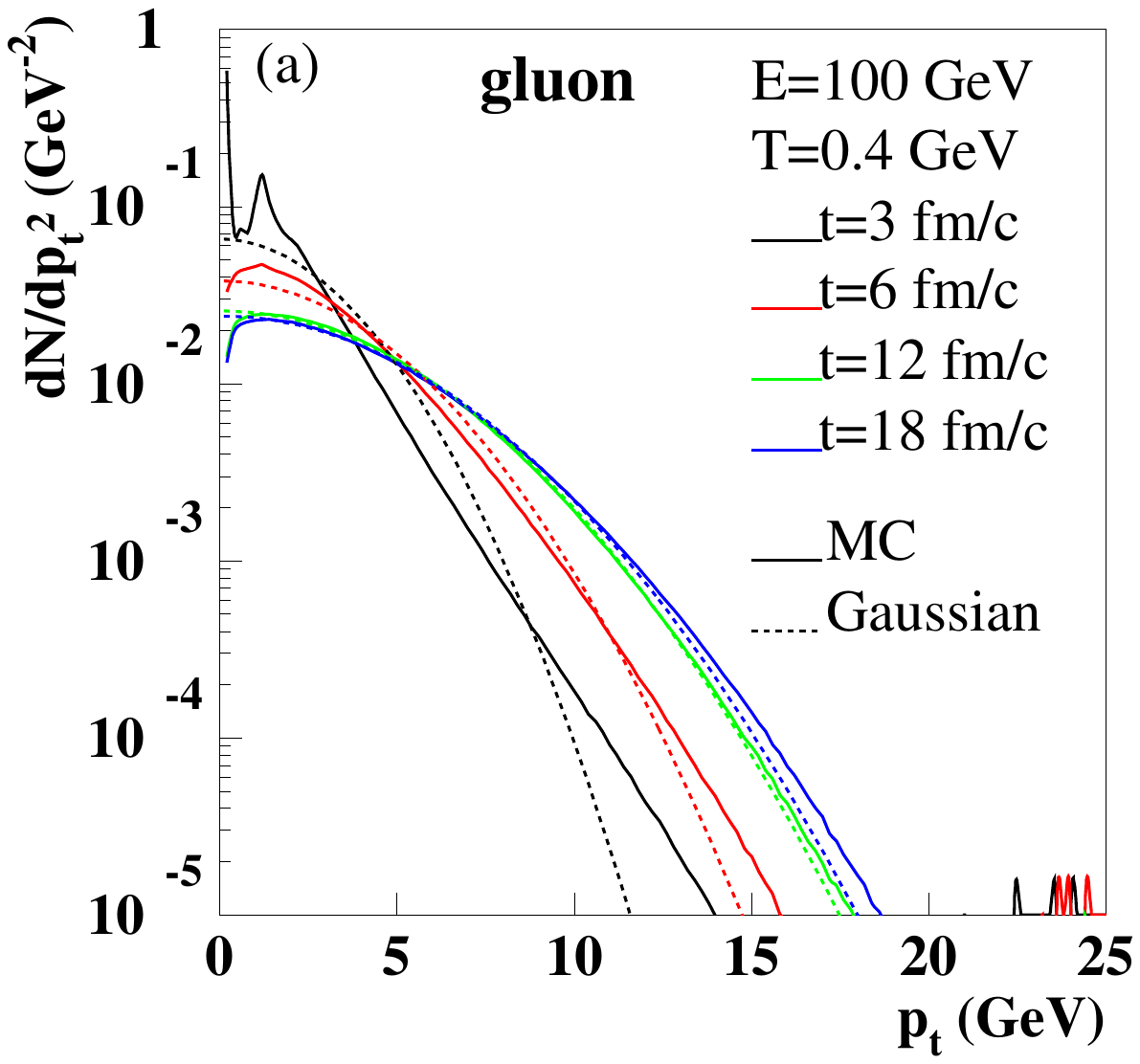}\\
\includegraphics[width=8.5cm,bb=15 150 585 687]{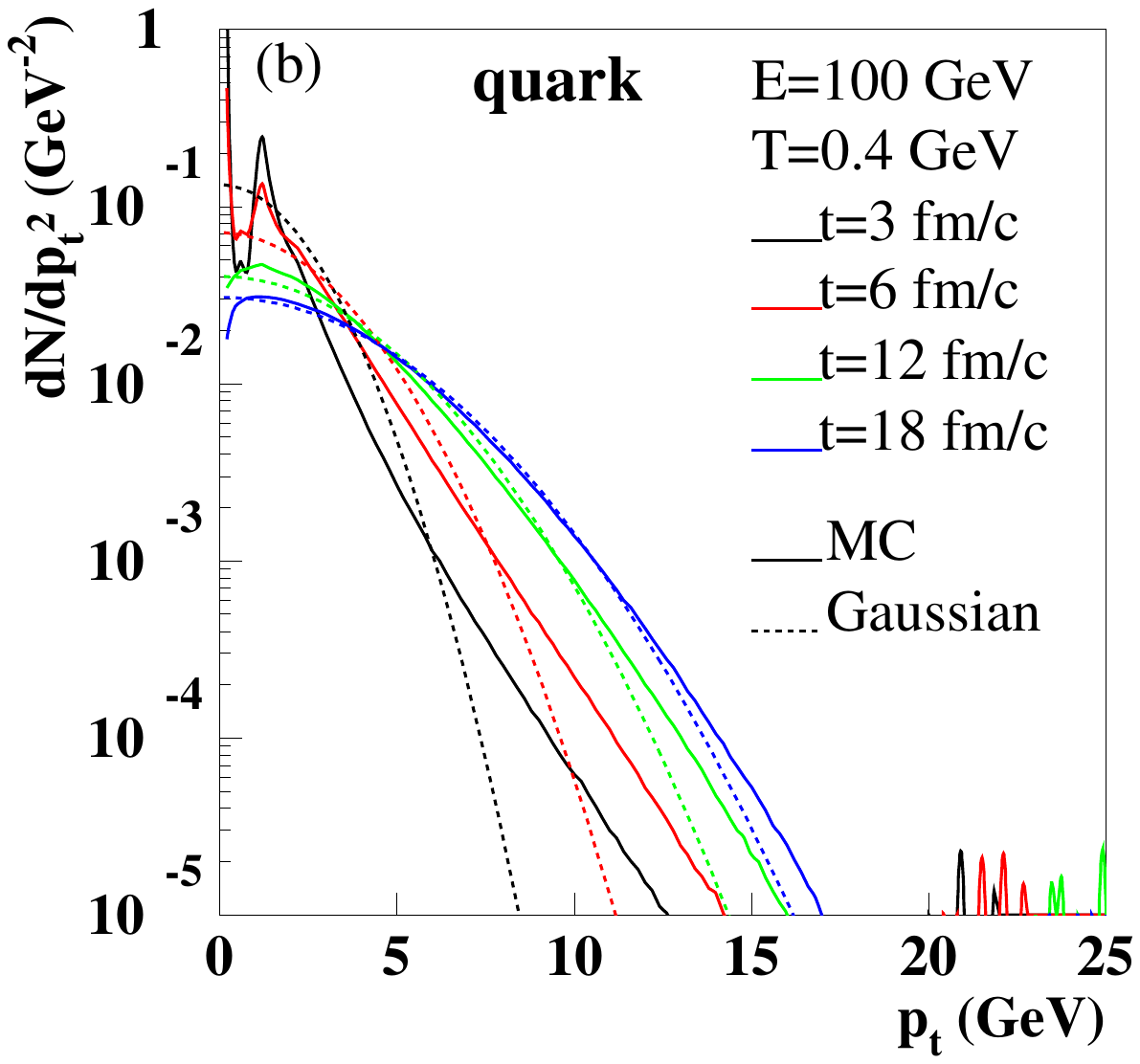}
\caption{(Color online) Transverse momentum distributions (solid lines) of the leading jet parton at different times for an initial (a) gluon or (b) quark propagating in a uniform and static medium at a temperature $T=400$ MeV. Gaussian fits are shown in dashed lines. }
\label{pt-broadening}
\end{figure}

We also show in Fig.~\ref{pt-broadening} the transverse momentum distributions  (solid lines) of the leading parton at different times for an initial gluon (upper panel)  or quark  (lower panel) with energy $E_0=100$ GeV propagating in a uniform medium at a temperature $T=400$ MeV. The transverse momentum broadening manifests in these distributions as the increase of the widths of distributions with time. Such an increase of the width gives rise to an approximately time-independent jet transport parameter $\hat q$ as shown in Fig.~\ref{qhat}. To illustrate the evolution of the transverse momentum distribution with time beyond the change of the width, we also compare each distribution with a Gaussian,
\begin{equation}
\label{ptfitfunc}
\frac{dN}{d p_T^2}=\frac{1}{\left\langle p_{T}^{2} \right\rangle } \, \exp (-\frac{p_{T}^{2}}{\left\langle p_{T}^{2} \right\rangle }),
\end{equation}
where $\langle p_{T}^{2} \rangle$ is the average transverse momentum squared from each corresponding transverse momentum distribution from LBT simulations. The $p_T$ distribution from a single scattering follows a power law $\sim 1/p_T^4$ at small angles $p_T^2\ll s^*/4$. At later times, the distribution deviates from the power law behavior due to multiple scattering.  According to Eqs.~(\ref{analytic_gammag}) and (\ref{analytic_gammaq}), the total scattering rate for a gluon (quark) is $\Gamma_{g} \, (\Gamma_q) \approx 1.53$ (0.68) 1/fm. The leading parton should experience several scatterings with medium partons after a few fm/$c$ of time. Its transverse momentum distribution will start to deviate from the power law and approach to a Gaussian according to the central limit theorem because the transverse momentum transfers during each scattering are independent of each other.  Since a gluon's scattering rate is more than twice that of a quark, its approach  a Gaussian form of the $p_T$ distribution is also twice sooner than a quark. In the high $p_T$ region, the power-law-like tail, however, will always remain due to a single hard or large-angle scattering. This evolution of the transverse momentum distribution of the leading parton is also shown in a recent perturbative calculation within a HTL pQCD approach \cite{deramo}.

\subsection{Elastic energy loss}

\begin{figure}
\includegraphics[width=8.5cm,bb=15 150 585 687]{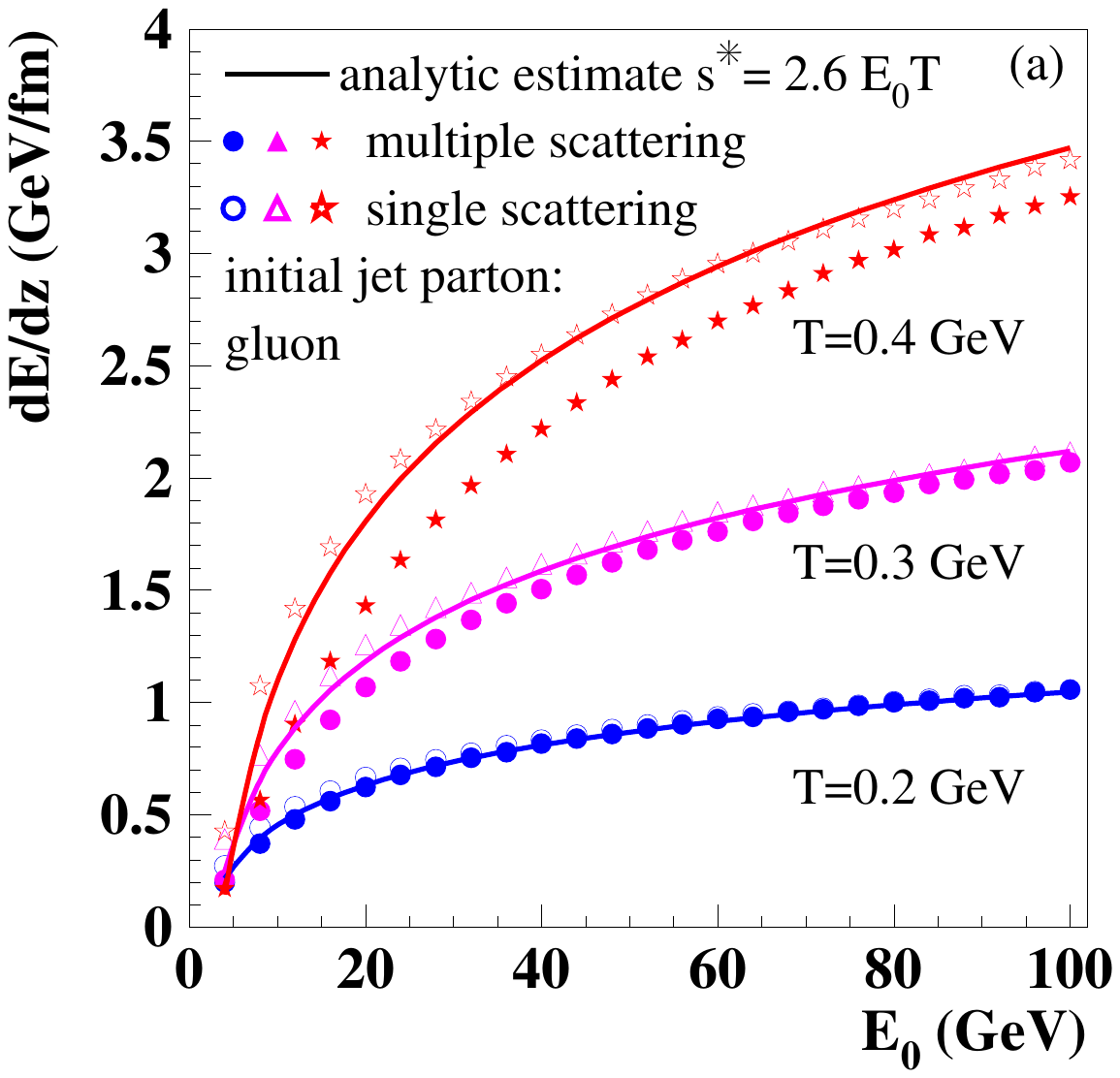}\\
\includegraphics[width=8.5cm,bb=15 150 585 687]{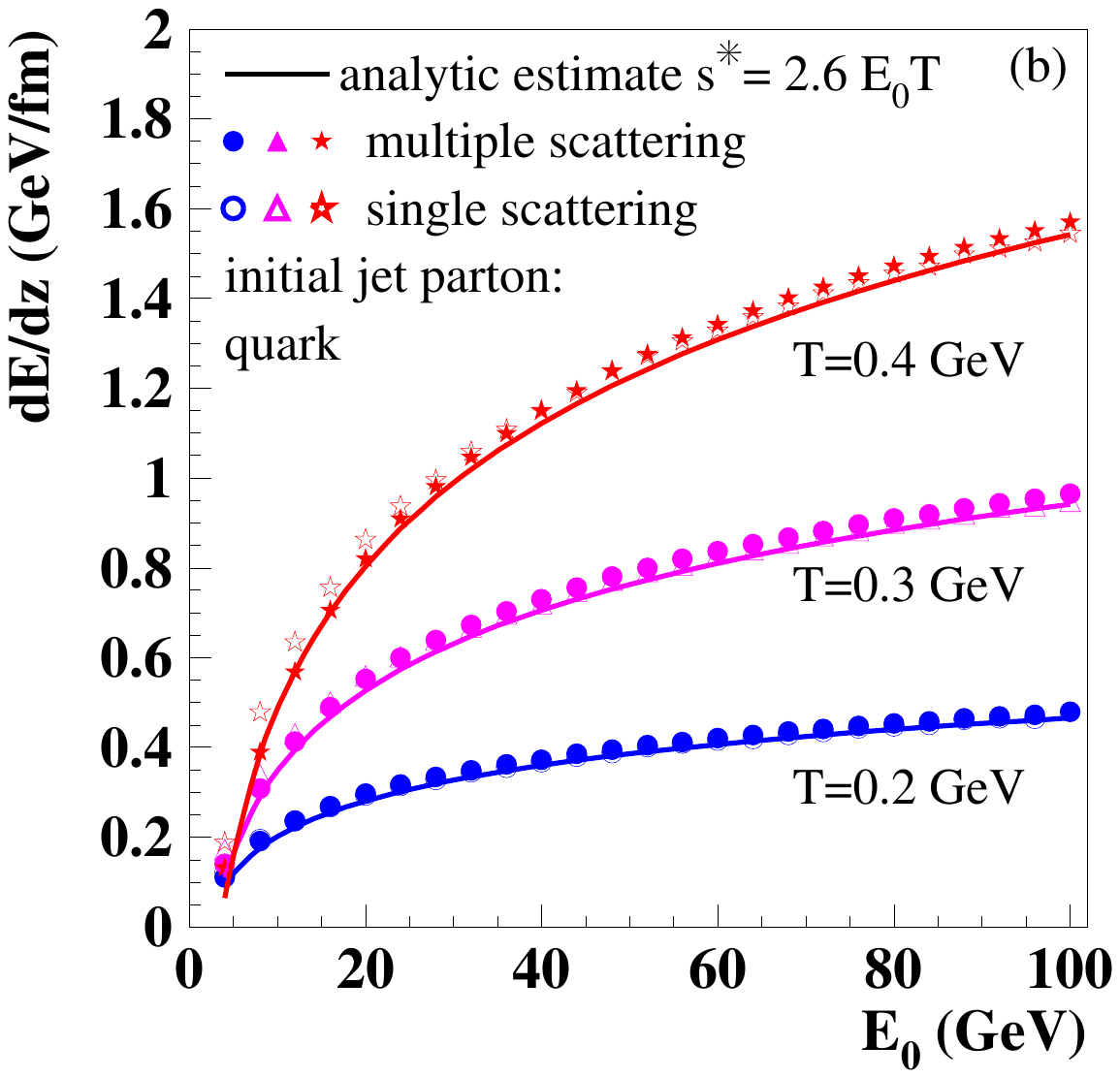}\\
\caption{(Color online) Elastic energy loss per unit length $dE/dz$ of (a) a gluon or (b) a quark in a uniform and static medium with a temperature $T=200$, 300 and 400 MeV as a function of the initial energy $E_0$ from LBT simulations of a single scattering (open symbols) or multiple scatterings ($L$=8fm) (solid symbols) as compared to analytic results (solid lines) with a small-angle approximation.}
\label{fig:elossE}
\end{figure}

Following the propagation of the leading parton, one can also calculate the elastic energy loss in the LBT model. Shown in Fig.~\ref{fig:elossE} is the calculated elastic energy loss per unit length  from LBT simulations as a function of the initial parton's energy for a gluon (upper panel) or a quark (lower panel) propagating in a uniform QGP medium with a length $L=8$ fm at different constant temperatures.  Also shown in the figure are the simulated elastic energy loss per unit mean-free-path length from a single scattering,
\begin{equation}
\label{eloss}
\frac{dE_{\rm el}^{a}}{d\lambda}=\langle \Gamma_a\nu \rangle,
\end{equation}
where $\nu$ is the energy transfer of the leading parton to a thermal medium parton which should depend on the energy $\omega$ of the thermal parton and the transverse momentum transfer squared $q_{\bot }^{2}$ in each scattering. Results from a single scattering are in good agreement with that from multiple scatterings over a long distance, especially for more energetic quarks at lower temperatures. The difference between results from a single and multiple scatterings becomes bigger for lower energy gluons at higher temperature, which can be understood by the energy dependence of the energy loss in a single scattering. Using the small-angle approximation for elastic scattering cross sections
in Eq.~(\ref{eq-small-el}) and $\nu \approx q_{\bot}^2 /{2\omega}$, the elastic energy loss per mean free path can be expressed as the product of the averaged transverse momentum transfer squared per scattering $\langle q_{\bot }^{2} \rangle\approx \mu_D^2\ln(s^*/4\mu_D^2)$ and the $1/\omega$-weighted scattering rate \cite{Wang:1996yf}:
\begin{equation}
\left\langle \frac{\Gamma_a}{2\omega} \right\rangle =C_a \frac{3\pi \alpha _{\rm s}^{2}}{2\mu _{D}^{2}} \, {{T}^{2}}.
\end{equation}
One obtains the analytic form of the elastic energy loss per unit mean free path,
\begin{equation}
\label{elossg}
\frac{dE_{\rm el}^a}{d\lambda}=C_a\frac{3\pi}{2} \alpha_{\rm s}^2 T^2 \ln (\frac{s^*}{4\mu _D^2}),
\end{equation}
under the small-angle scattering approximation. This analytic result agrees with the full HTL pQCD result \cite{Thoma1,Thoma2}  in the leading logarithmic approximation and has a logarithmic dependence on the initial parton energy. As shown in Fig.~\ref{fig:elossE} (solid lines), this analytic result under small-angle scattering approximation agrees well with LBT simulations of a single scattering.

\begin{figure}
\includegraphics[width=8.5cm,bb=15 150 585 687]{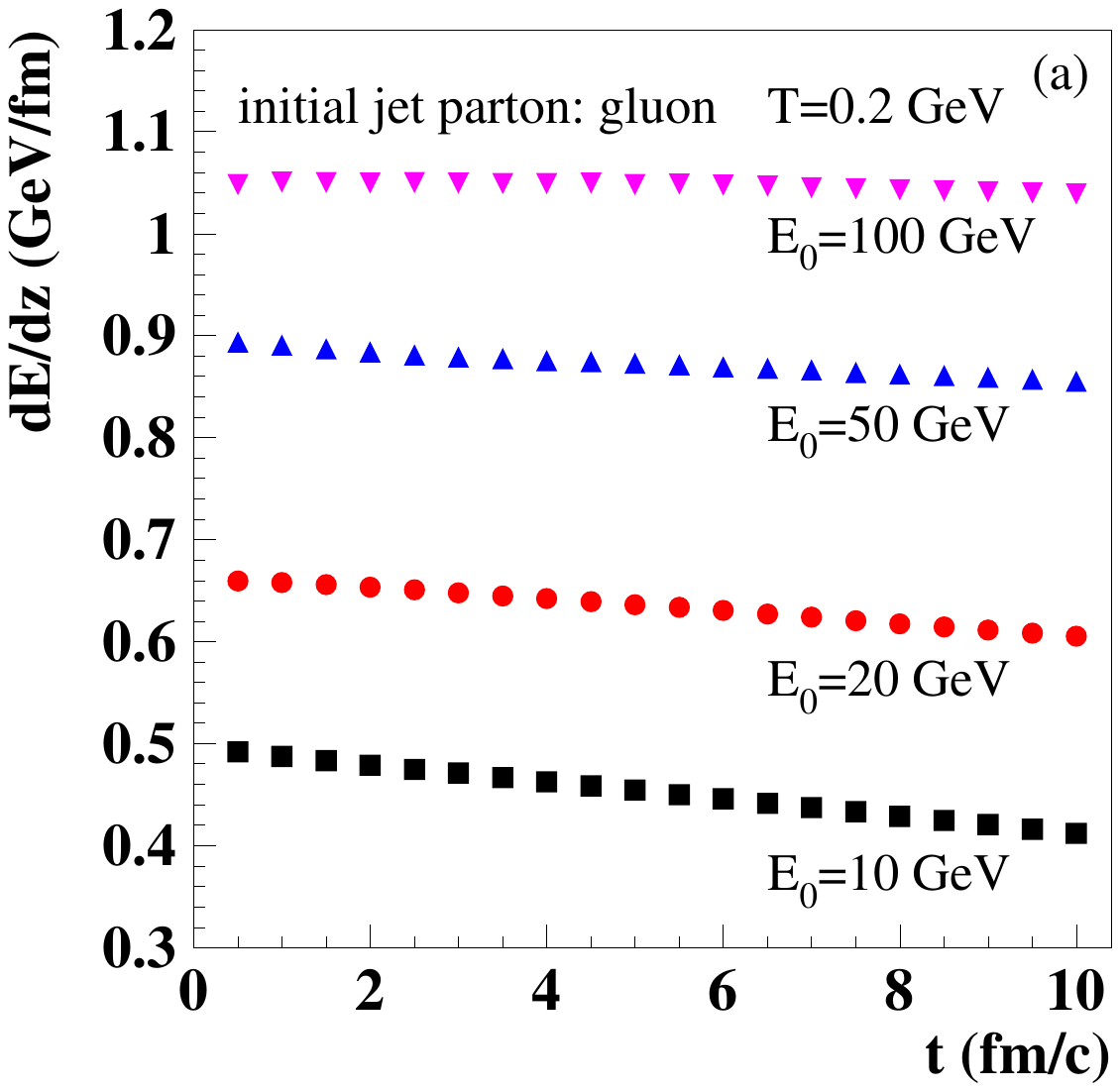}\\
\includegraphics[width=8.5cm,bb=15 150 585 687]{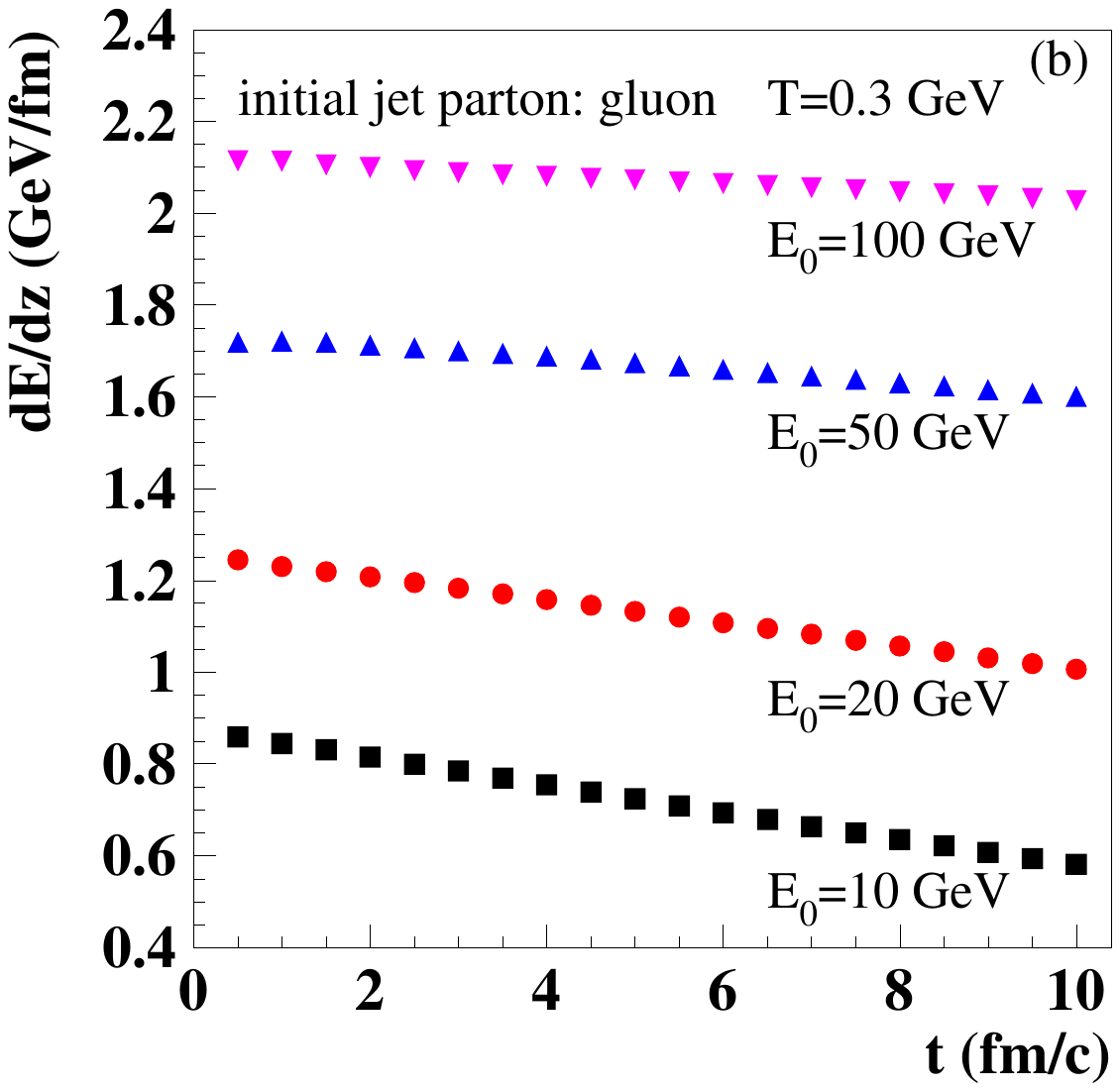}\\
\caption{(Color online) Energy loss per unit path length $dE/dz$ of a gluon with different initial energies in a uniform medium with a constant temperature (a) $T=200$ MeV and (b) $T=300$ MeV as a function of the propagation time.}
\label{fig:elosstimeg}
\end{figure}

\begin{figure}
\includegraphics[width=8.5cm,bb=15 150 585 687]{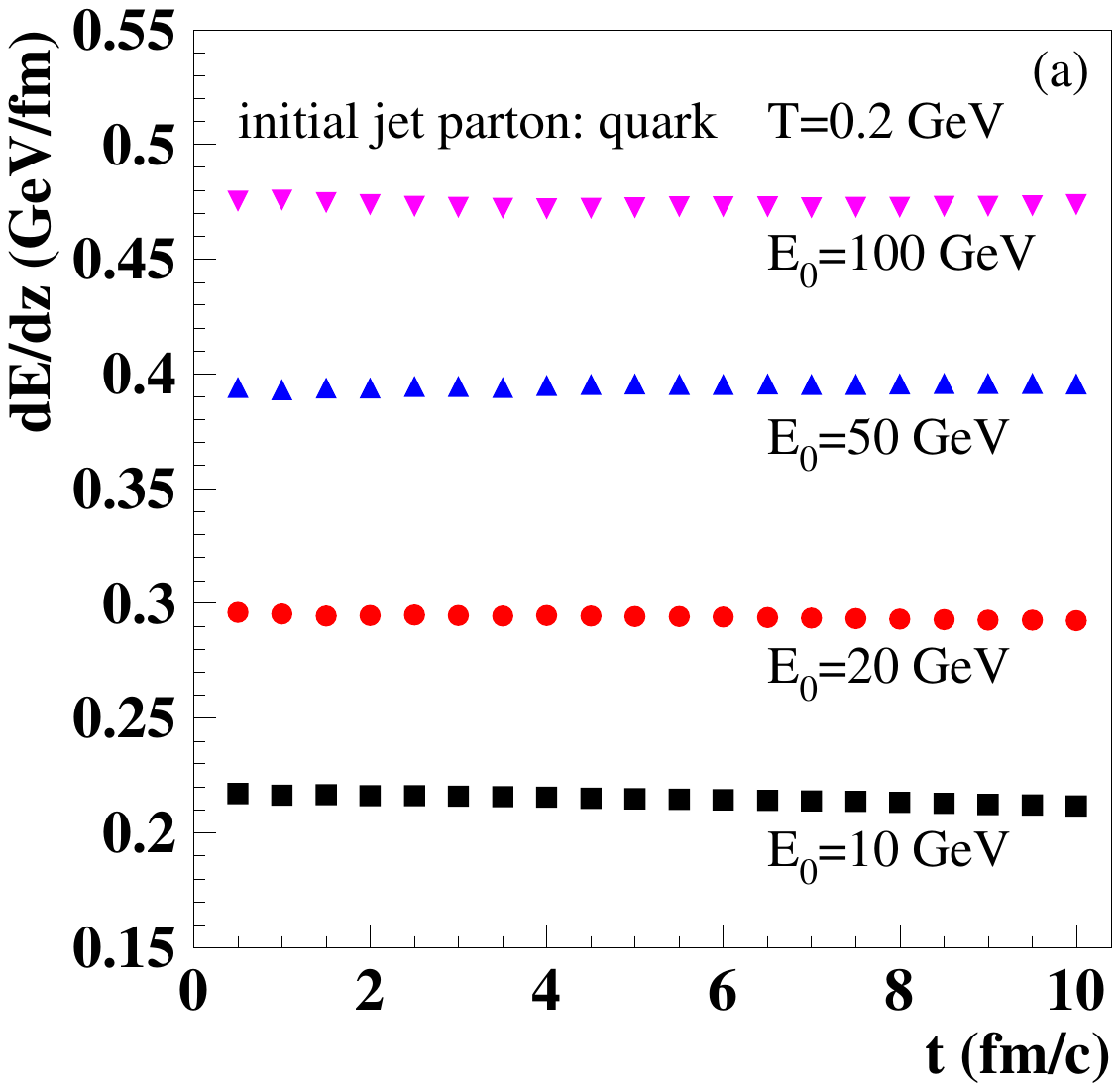}\\
\includegraphics[width=8.5cm,bb=15 150 585 687]{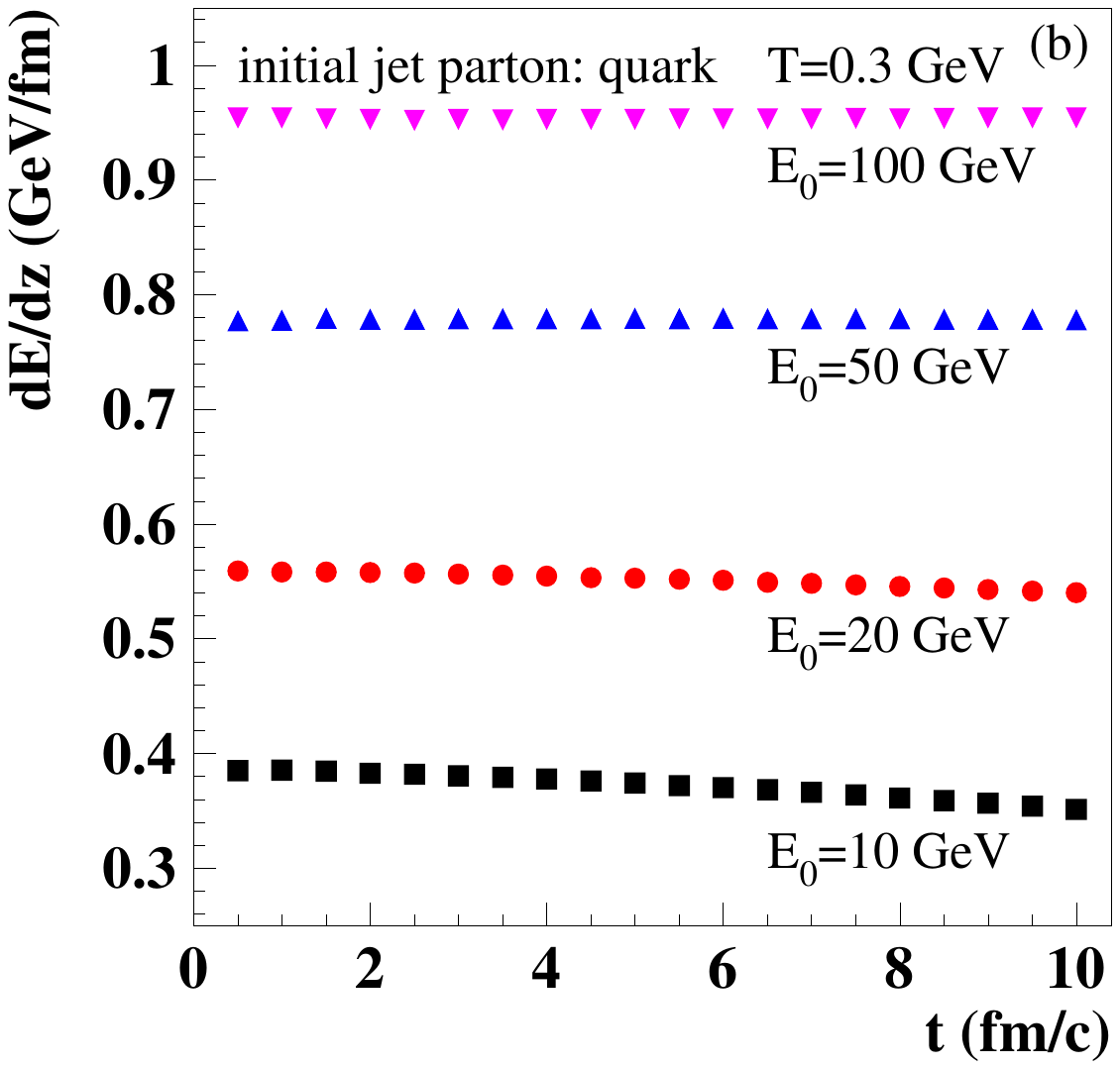}\\
\caption{(Color online) The same as Fig.~\ref{fig:elosstimeg} for a quark.}
\label{fig:elosstimeq}
\end{figure}

The logarithmic energy dependence of the elastic energy loss can also explain the difference between single and multiple scatterings in Fig.~\ref{fig:elossE}. During multiple scatterings in the LBT model, energy and momentum is conserved during each scattering. The leading parton will lose its energy along its propagation path. The reduced energy of the leading parton will then lead to smaller elastic energy loss for subsequent scatterings according to the logarithmic energy dependence. This will lead to an overall reduction of the averaged energy loss per unit distance during multiple scattering over a long distance. This reduction on the averaged energy loss per unit distance is more significant when the total energy loss $\Delta E=L dE/dz$ is comparable to parton's initial energy. Since a gluon's energy loss is $9/4$ times larger than that of a quark, this degradation of energy and the energy loss is more significant for a gluon with lower energies in a QGP medium at higher temperatures as shown in Fig.~\ref{fig:elossE}. To illustrate this degradation of energy loss over time, we show in
Figs.~\ref{fig:elosstimeg} and \ref{fig:elosstimeq} the averaged elastic energy loss per unit distance as a function of propagation time for different initial parton energies and medium temperatures. It is clear that the degradation of parton energy loss with time is the strongest for a gluon with the lowest initial energy and in a medium at the highest temperature. For an initial high energy quark, the energy loss remains almost a constant even at a later time.

\begin{figure}
\includegraphics[width=8.5cm,bb=15 150 585 687]{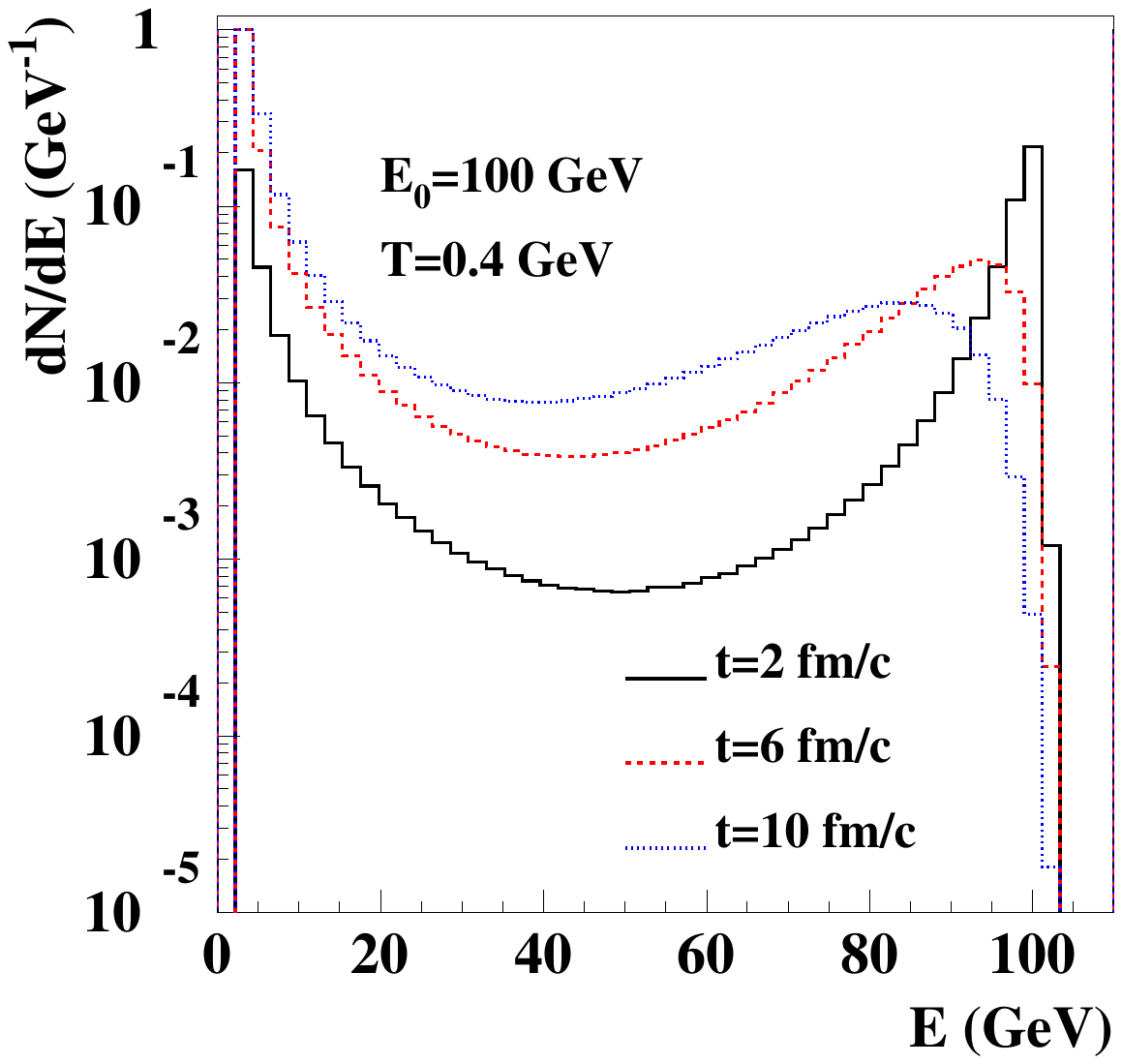}
\caption{(Color online) Energy spectra of leading partons and jet-induced medium partons (thermal recoil  partons minus ``negative" partons) at different times during the propagation of an initial gluon with energy $E_0=100$ GeV in a uniform QGP medium at a constant temperature $T=400$ MeV.}
\label{spectra-time}
\end{figure}

In the LBT model, energy and momentum are conserved in each parton-medium scattering (with the energy and momentum of ``negative" partons subtracted from the sum) along the path of the parton propagation. The energy lost by the leading parton as shown in the above calculations will be carried by thermal recoil   partons, minus that of ``negative" partons, in the form of jet-induced medium excitations. To illustrate the transfer of energy from the leading partons to jet-induced medium excitations, we plot in Fig.~\ref{spectra-time} the energy spectra of all partons including both leading and the thermal recoil   partons (with the ``negative" partons subtracted) at different times ($t=2$, 6, and 10 fm/$c$) during the propagation of an initial gluon with energy $E_0=100$ GeV in a uniform QGP medium at a constant temperature $T=400$ MeV. At the beginning of the parton propagation, leading partons peak around the initial energy $E_0$ with power law tails towards small values due to the dominant $t$-channel scattering, while thermal recoil   partons (minus ``negative" partons) peak around their thermal energy $\sim T+\mu_D^2/T$ with a power law tail. At later times, leading partons continue to lose energy due to multiple scattering and peak at smaller energies $\sim E_0-t dE/dz$ with increasingly broadened distributions. The number of jet-induced medium partons (thermal recoil   partons minus ``negative" partons), on the other hand, will continue to increase with softened spectra which resemble a thermal distribution at later times. In principle, the spectra of leading and jet-induced medium partons will eventually merge and approach a thermal spectrum as a result of the equilibration. But this limit is beyond the validity of the LBT model.

\section{Jet-induced Medium excitations}

During the parton propagation in a QGP medium, the energy and momentum deposited in the medium should dissipate with time and propagate like a sound wave. Since the leading massless parton travels at the speed of light which is much faster than the maximum sound velocity $c_s=1/\sqrt{3}$ in an ideal QGP, the sound wave induced by the propagating parton becomes supersonic. It can form a Mach-cone shock wave as shown in many recent studies \cite{CasalderreySolana:2004qm,Stoecker:2004qu,Chaudhuri:2005vc,Betz:2008ka,Qin:2009uh,Neufeld:2009ep,Chesler:2007sv,Gubser:2007ga}. Simulations of jet propagation within parton transport models also show similar Mach-cone-like  features \cite{Bouras:2012mh,Li:2010ts}. Studies of such jet-induced medium excitations can help us to understand the experimental data on dissipation of energy lost by quenched jets \cite{Tachibana:2014lja} and extract bulk medium properties such as sound velocity or the equation of state  (EoS) of the QGP medium. Since this article is devoted to a general description and test of the LBT model in a uniform medium, we will focus on the effect of dynamical scatterings of propagating partons in a uniform medium. This is similar to the effect of viscosity on the propagation of shockwaves in a QGP medium \cite{Bouras:2012mh,Bouras:2009nn,Bouras:2010hm,Bouras:2014rea}.  We will discuss the propagation of jet-induced Mach cones in an expanding QGP medium in the future when we carry out realistic studies of jet-medium interaction in high-energy heavy-ion collisions.

\subsection{Diffused Mach cone}

To study jet transport and jet-induced medium excitations at the same time in a QGP medium, both of the outgoing partons in each $2\rightarrow 2$ scattering are tracked in LBT Monte Carlo simulations. They are allowed to undergo further propagation and scattering in the medium.  
Of the two outgoing partons, 
the more energetic one is designated as the leading parton for further propagation and scattering. The flavor of the outgoing leading parton can be different from incident parton though this is quite rare since the scattering is dominated by small-angle $t$-channel processes. The less energetic or soft parton from each parton-medium scattering is considered as the recoil medium parton that is also allowed to further propagate and interact with the medium. In addition, the initial parton from the thermal medium participating in the scattering is also tracked and allowed to further propagate in the LBT model whose four-momentum is subtracted from the final result to account for the back-reaction in the Boltzmann transport equation. These thermal partons from the back-reaction are denoted as ``negative" partons in our study and are included in all final results on parton spectra and jet reconstruction. In general, we refer to the thermal recoil   and ``negative" partons as jet-induced thermal partons.

Since jet-induced thermal partons are allowed to propagate in the QGP medium according to the Boltzmann transport equation in the LBT model, we can study the linear response of the medium to jet-medium interaction along the path of a jet's propagation. Interactions among jet-induced medium partons are neglected. Such an approximation of linear response is valid when the jet-induced medium excitation is much smaller than the thermal background, $\delta f\ll f$.

\begin{figure}
\includegraphics[width=9.0cm]{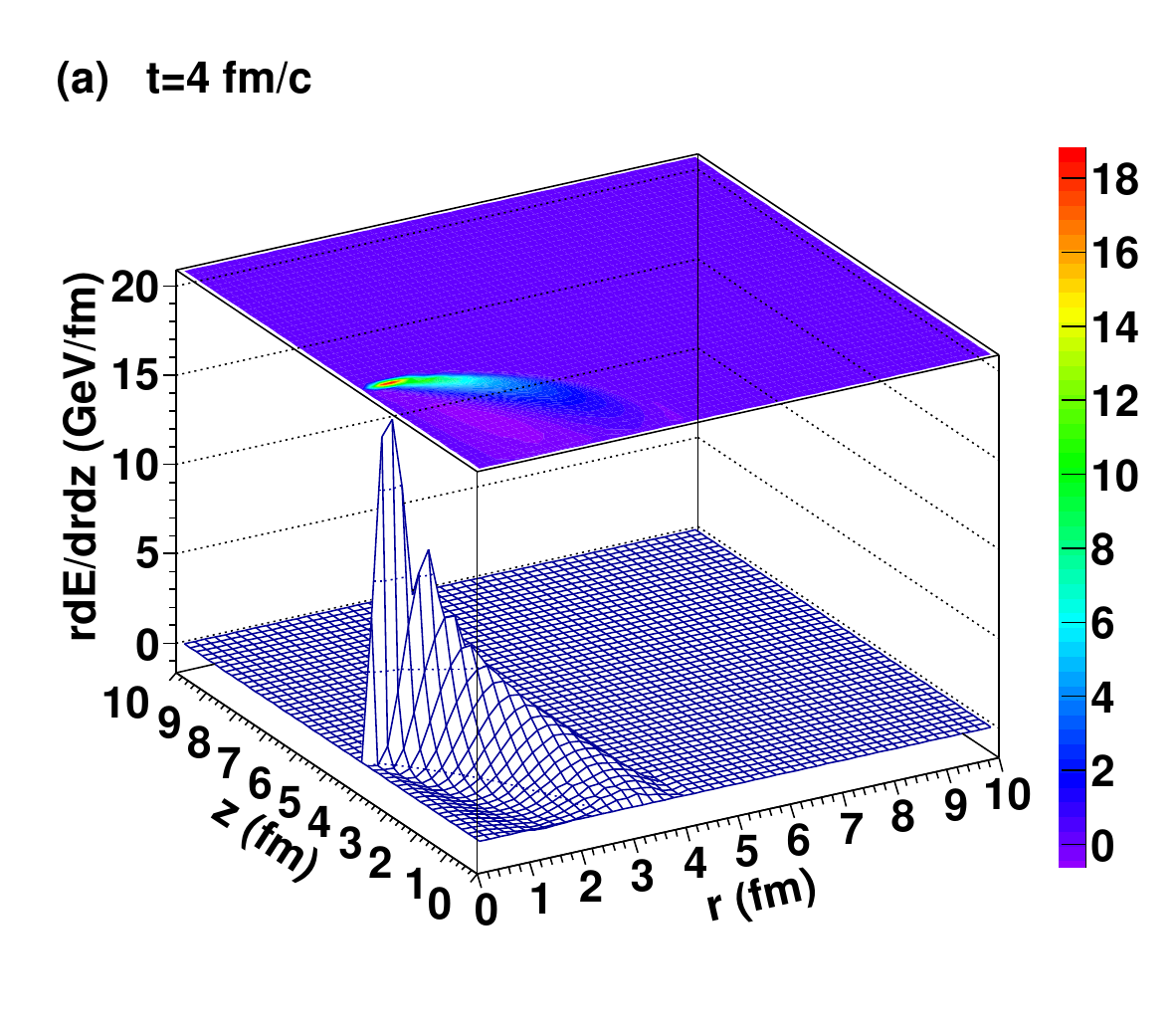}\\
\vspace{-0.4 cm}
\includegraphics[width=9.0cm]{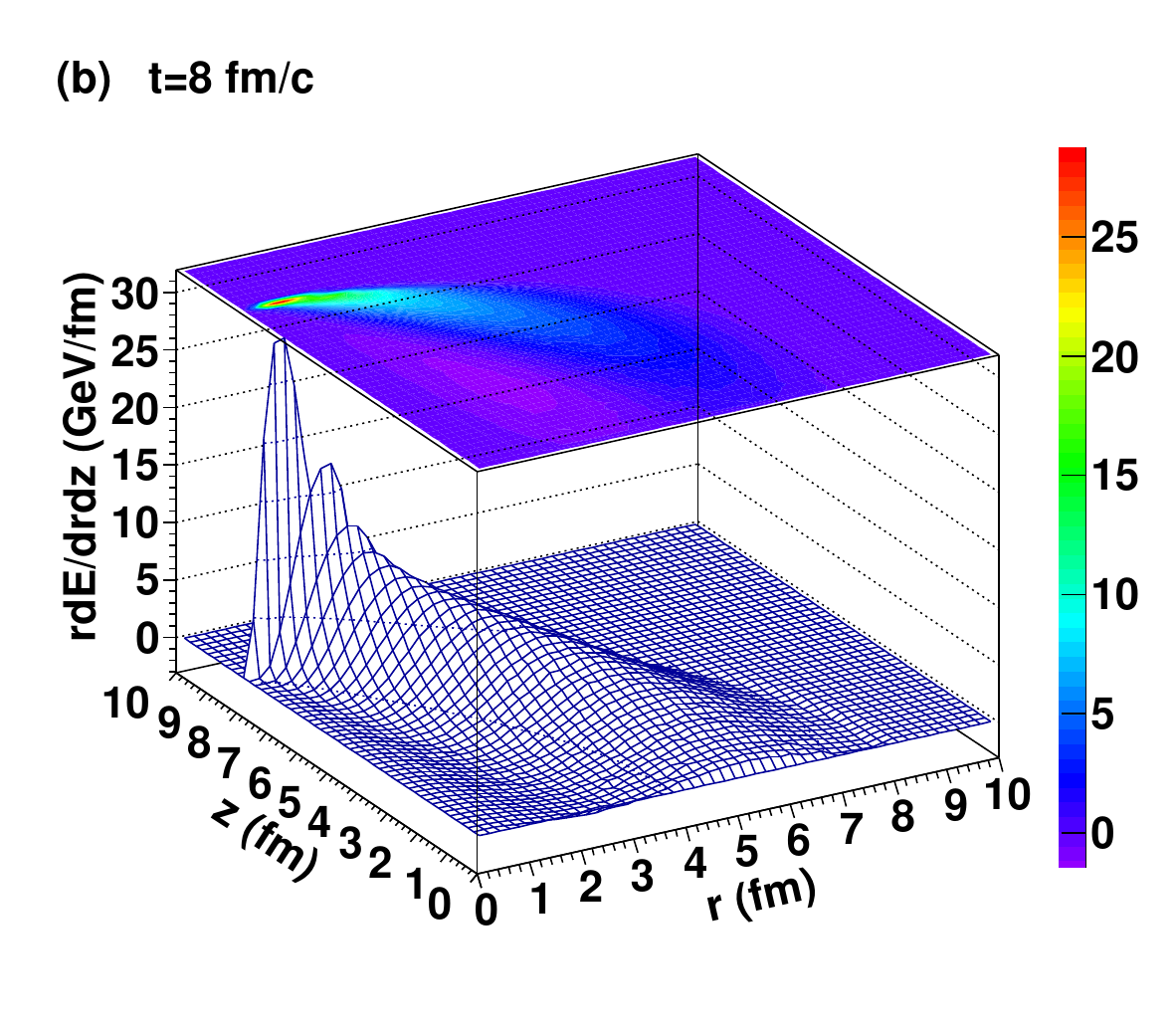}
\caption{(Color online) Energy density distribution of the jet-induced medium excitation by a gluon with an initial energy $E_0=100$ GeV along ($z$) and transverse ($r$) to the initial parton direction after (a) 4  and (b) 8 fm of propagation in a uniform QGP medium at a constant temperature $T=400$ MeV. The jet parton starts propagating along the $z$ direction at $z=0$ and $r=0$.}
\label{fig:g-cone}
\end{figure}

The energy and momentum lost to the medium by the propagating parton will act like a source for jet-induced medium excitation through the propagation and transport of the jet-induced medium partons.  Shown in Fig.~\ref{fig:g-cone} are two-dimensional  and contour energy density distributions of the medium excitation induced by a propagating gluon with an initial energy $E_0=100$ GeV along ($z$) and transverse ($r$) to the initial parton direction after 4 fm/$c$ (upper panel) and 8 fm/$c$ (lower panel) of propagation in a uniform QGP medium at a constant temperature $T=400$ MeV.  The jet parton starts propagating along the $z$ direction at $z=0$ and $r=0$. Note that in each event of parton propagation there are only a limited number of jet-induced medium partons. Figure \ref{fig:g-cone} shows the energy density distributions of jet-induced medium partons or medium excitation averaged over many events.  The number of medium recoil and ``negative" partons increases separately with time in LBT Monte Carlo simulations. The net energy density, however, remains approximately constant because of energy-momentum conservation and the near constant energy-momentum transfer rate via parton-medium interaction. One can see clearly the formation and propagation of a Mach-cone-like shock wave trailing the leading parton. The edge of the shock wave travels at a speed limited by the velocity of light while the peak of the shock wave propagates at the effective sound velocity in the medium. The shock wave is significantly diffused during its propagation because of the dissipation due to finite values of viscosities as a result of parton-parton collisions as implemented in the LBT model.  The energy density of jet-induced medium partons is negative behind the propagating parton. This is the diffusion wake induced by the jet-medium interaction which essentially depletes the thermal medium parton density behind the path of the propagating parton.

The diffused shock wave in the shape of an elliptic paraboloid induced by jet-medium interaction within the LBT model as shown in Fig.~\ref{fig:g-cone} is sharply different from the Mach-cone shape given by other linear response theories \cite{Neufeld:2009ep}, hydrodynamic simulations \cite{Chaudhuri:2005vc,Betz:2008ka,Qin:2009uh} or anti-de Sitter/Conformal field theory (AdS/CFT) correspondence studies \cite{Chesler:2007sv,Gubser:2007ga} with a source term that travels along a given direction. The leading parton in the LBT Monte Carlo simulations, however, does not travel along a fixed direction. It receives transverse momentum transfers in random directions and accumulates transverse momentum broadening along its propagation path. Such momentum broadening amounts to a smearing in the direction of the source term for jet-induced medium excitation and the width of the smearing increases with the propagation time. This eventually leads to the elliptic paraboloid shape of the shock wave instead of a sharp Mach-cone.

\begin{figure}
\includegraphics[width=8.0cm,bb=15 150 585 687]{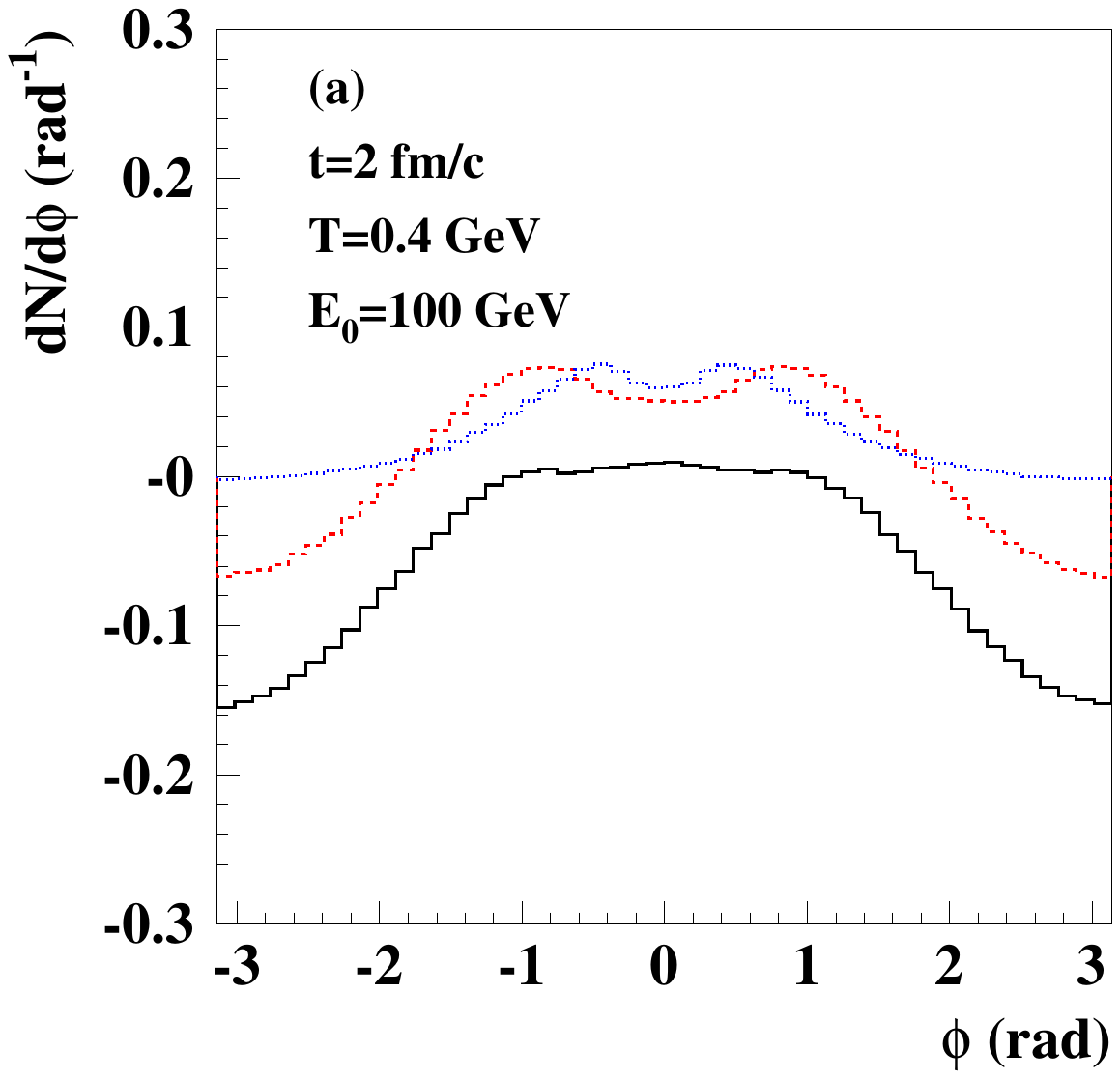}\\
\vspace{-0.5cm}
\includegraphics[width=8.0cm,bb=15 150 585 687]{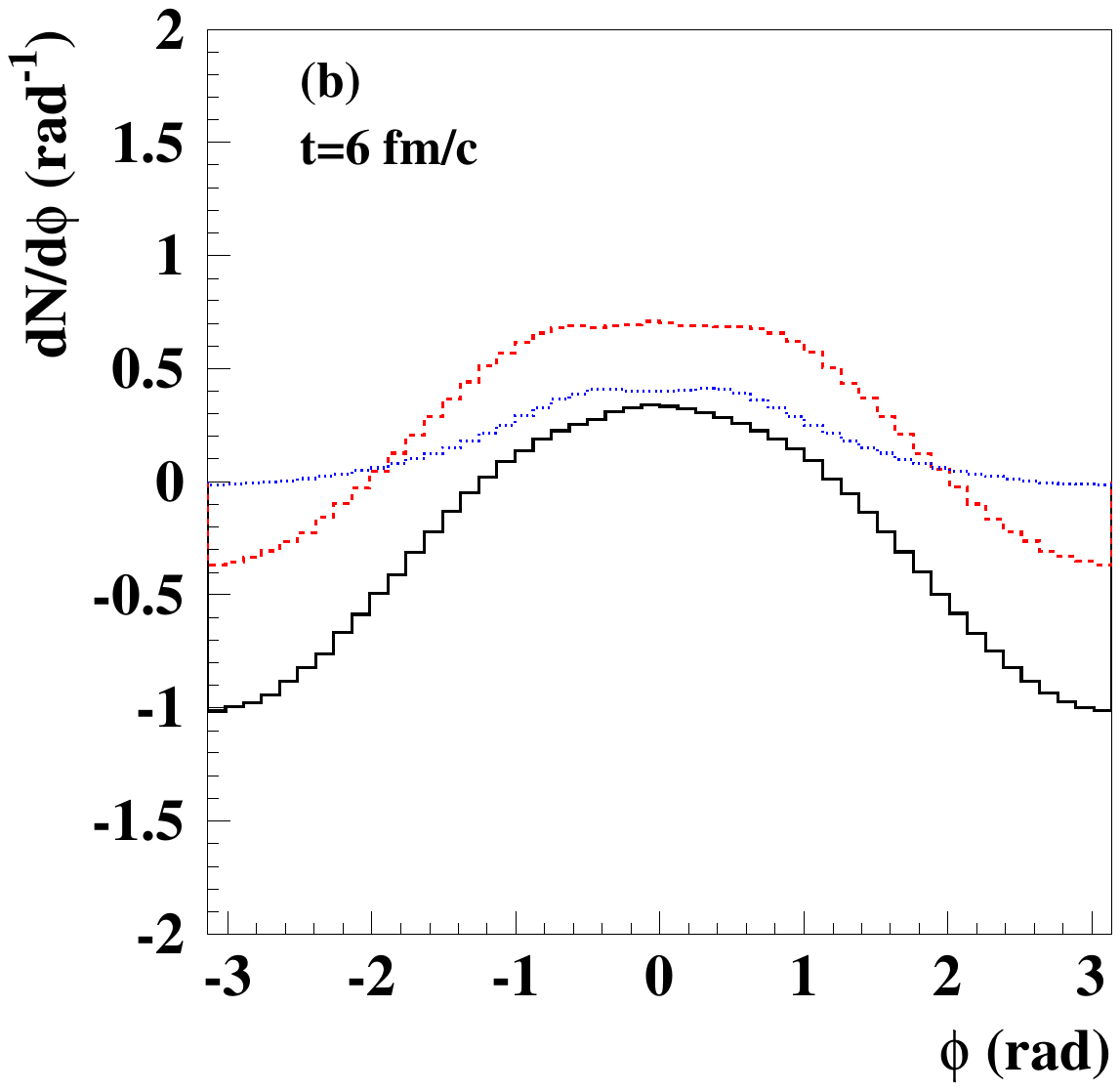}\\
\vspace{-0.5cm}
\includegraphics[width=8.0cm,bb=15 150 585 687]{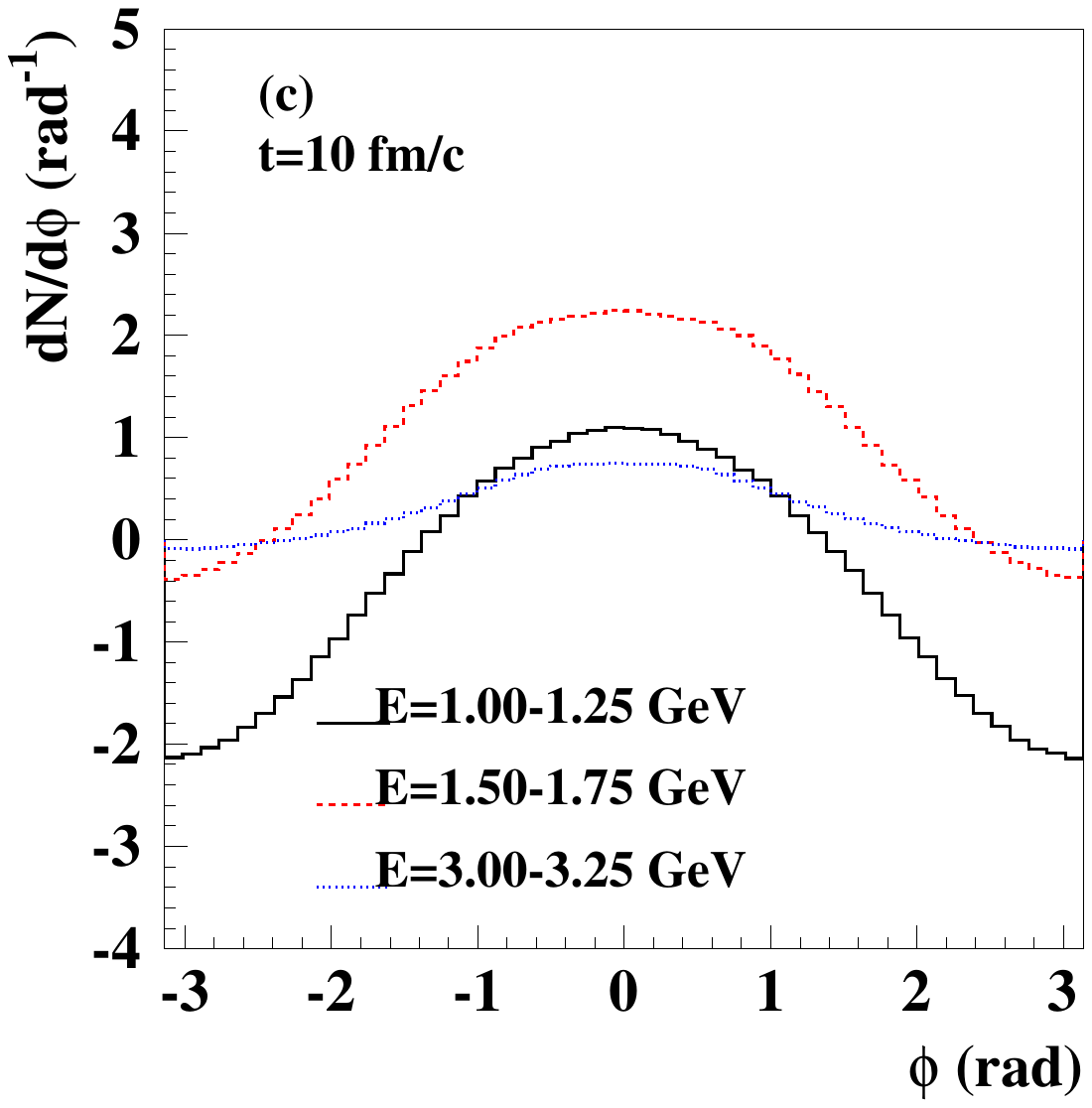}\\
\vspace{-0.5cm}
\caption{(Color online) Angular distributions of medium partons with energy $E=1.0-1.25$ (solid), 1.5-1.75 (dashed), 3.0-3.25 GeV (dotted) induced by a gluon with an initial energy $E_0=100$ GeV after (a) $t=2$, (b) 6  and  (c) 10 fm/$c$ of propagation in a uniform QGP medium at a constant temperature $T=400$ MeV.}
\label{fig:g-angle100}
\end{figure}

\subsection{Azimuthal diffusion}

To illustrate consequences of the transverse momentum broadening of leading partons on the final spectra of jet-induced medium partons, we show in Fig.~\ref{fig:g-angle100}  angular distributions of jet-induced medium partons with different ranges of energy relative to the initial direction of a gluon with energy $E_0=100$ GeV after  $t=2$ (upper panel), 6 (middle panel), and 10  fm/$c$ (lower panel) of propagation in a uniform medium at a constant temperature $T=400$ MeV.   In a uniform medium, event-averaged parton distributions should be azimuthal symmetric with respect to the initial direction of the propagating parton.  For the convenience of future study of azimuthal distributions in a cylindrical frame along the beam direction in high-energy heavy-ion collisions, we project the angular distribution onto a plane aligned with the initial propagating parton,
\begin{equation}
\frac{dN}{d\phi}=\int d\theta d\varphi \frac{dN}{d\theta d\varphi} \delta\left( \phi - \arctan[\tan \theta \cos\varphi]\right),
\end{equation}
where $\theta$ and $\varphi$ are the polar and azimuthal angles, respectively, relative to the initial jet parton direction.  We refer to the angle $\phi$ as the projected azimuthal angle and the distribution as the projected azimuthal distribution. Soft jet-induced medium partons ($E=0-1$ GeV) are mostly ``negative" partons from the the back-reaction whose contributions to the spectra are negative according to our subtraction scheme. These ``negative" partons have a broad angular distribution that does not change much over time.  Contributions to more energetic jet-induced medium partons from ``negative" partons decrease with time, especially along the initial direction of the propagating parton, and thermal recoil  partons dominate. Intermediate energy thermal recoil  partons during the early time have an angular distribution similar to that of a single scattering as shown in Figs.~\ref{fig-theta-gluon} and \ref{fig-theta-quark} that has two peaks away from the initial direction. At later times, however, the double-peak structure disappears due to transverse momentum broadening of the leading parton and diffusion of thermal recoil  partons due to multiple scatterings.  The effect of the transverse momentum broadening on the angular distribution of thermal recoil   partons is stronger for less energetic initial partons at later times, especially when the total energy loss becomes sizable relative to the initial parton energy.


\section{Reconstructed jets in medium}

In high-energy heavy-ion collision experiments, reconstructed jets have become a powerful tool for the study of jet quenching \cite{Aad:2010bu,Chatrchyan:2011sx,Chatrchyan:2012gt,Qin:2010mn,Young:2011qx,He:2011pd,Renk:2012cx,Zapp:2012ak}.  These jets are collimated clusters of hadrons and are reconstructed from the calorimetric energy of final hadrons within a jet cone,
\begin{equation}
\sqrt{(\eta-\eta_J)^2+(\phi-\phi_J)^2} \le R,
\end{equation}
using a jet finding algorithm \cite{Cacciari:2011ma}, where $\eta$ ($\eta_J$) and $\phi$ ($\phi_J$) are the pseudorapidity and azimuthal angle of the final hadron (jet), respectively. In heavy-ion collisions, one should also subtract the background energy within the jet cone from hadrons in the underlying events. Such background subtraction is extremely nontrivial and may affect the final jet energy and jet spectra.

If we neglect nonperturbative effects of parton hadronization, we can approximate final hadronic jets with partonic jets reconstructed from final partons in  theoretical simulations. In our LBT Monte Carlo simulations, we assume a perfect subtraction of the background from underlying events if the initial propagating parton is not present. We achieve this by including both leading partons as well as jet-induced medium partons for the jet reconstruction. We use a modified version of the  FASTJET  code \cite{Cacciari:2011ma} with the anti-$k_T$ algorithm for jet reconstruction in our study within the LBT model in which a ``negative" parton is treated as a normal one when its pseudorapidity and azimuthal angle are calculated, but its energy and momentum are subtracted from the total energy and momentum of the reconstructed jet in each iteration of the jet-finding algorithm. We only consider the leading jet per event through the FASTJET jet finding package with the given algorithm. 

\subsection{Jet energy loss and $p_T$ broadening}

\begin{figure}
\includegraphics[width=8.5cm,bb=15 150 585 687]{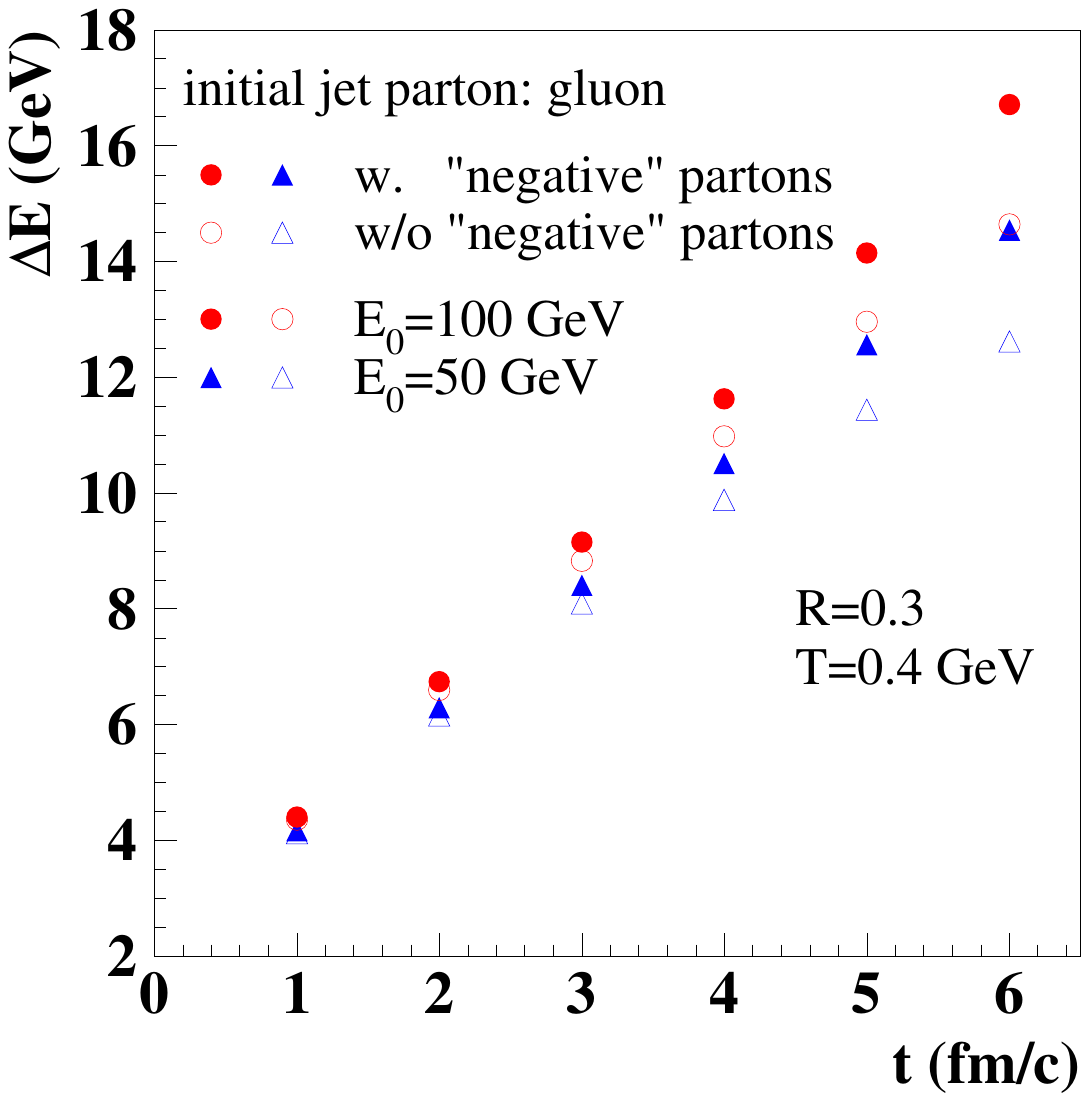}\\
\caption{(Color online) Jet energy loss in a uniform QGP medium at a temperature $T=400$ MeV as a function of propagation time of an initial gluon with energy $E_0=50$ and 100 GeV. Jets are reconstructed with all partons (leading and jet-induced medium partons) (solid symbols) or without ``negative" partons (open symbols). }
\label{jet-eloss}
\end{figure}

During the propagation of a fast parton, it undergoes multiple scatterings with medium partons in the LBT model. The leading parton loses energy and experiences transverse momentum broadening. The lost energy and momentum transfer are then carried by jet-induced medium partons which will propagate and go through further scattering with the medium leading to jet-induced medium excitations. Some fraction of the energy and momentum carried by jet-induced medium partons will still be confined within the jet cone. However, an increasing fraction will be transported outside the jet cone via large-angle scattering and diffusion of jet-induced medium partons over the course of the parton propagation. This will lead to a reduction of the energy of the reconstructed jet. Shown in Fig.~\ref{jet-eloss} is the energy loss of reconstructed jets with a cone size $R=0.3$ as a function of time for an initial gluon with energy $E_0=50$ and 100 GeV in a uniform QGP medium with a constant temperature $T=400$ MeV. The solid symbols represent the energy loss of jets reconstructed with all of jet-induced medium partons while the open symbols represent jets without ``negative" partons. The jet energy loss increases linearly with time or distance of propagation similarly to the elastic energy loss of a propagating parton.

Th effect of ``negative" partons from the diffusion wake is negligible during the early stage of the parton propagation because there are very few of them.  At the later stage, as the number of these ``negative" partons becomes large, they deplete significantly the thermal medium behind the propagating parton and effectively modify the background underlying the jet. When such modification of the underlying background is taken into account via the subtraction of ``negative" partons from the jet cone, the effective jet energy loss becomes bigger as shown in Fig.~\ref{jet-eloss}. Without subtraction of ``negative" partons, the jet energy loss tends to saturate at later times. The linear time dependence of the total effective jet energy loss is restored only after the energy of ``negative" partons is subtracted.

\begin{figure}
\includegraphics[width=8.5cm,bb=15 150 585 687]{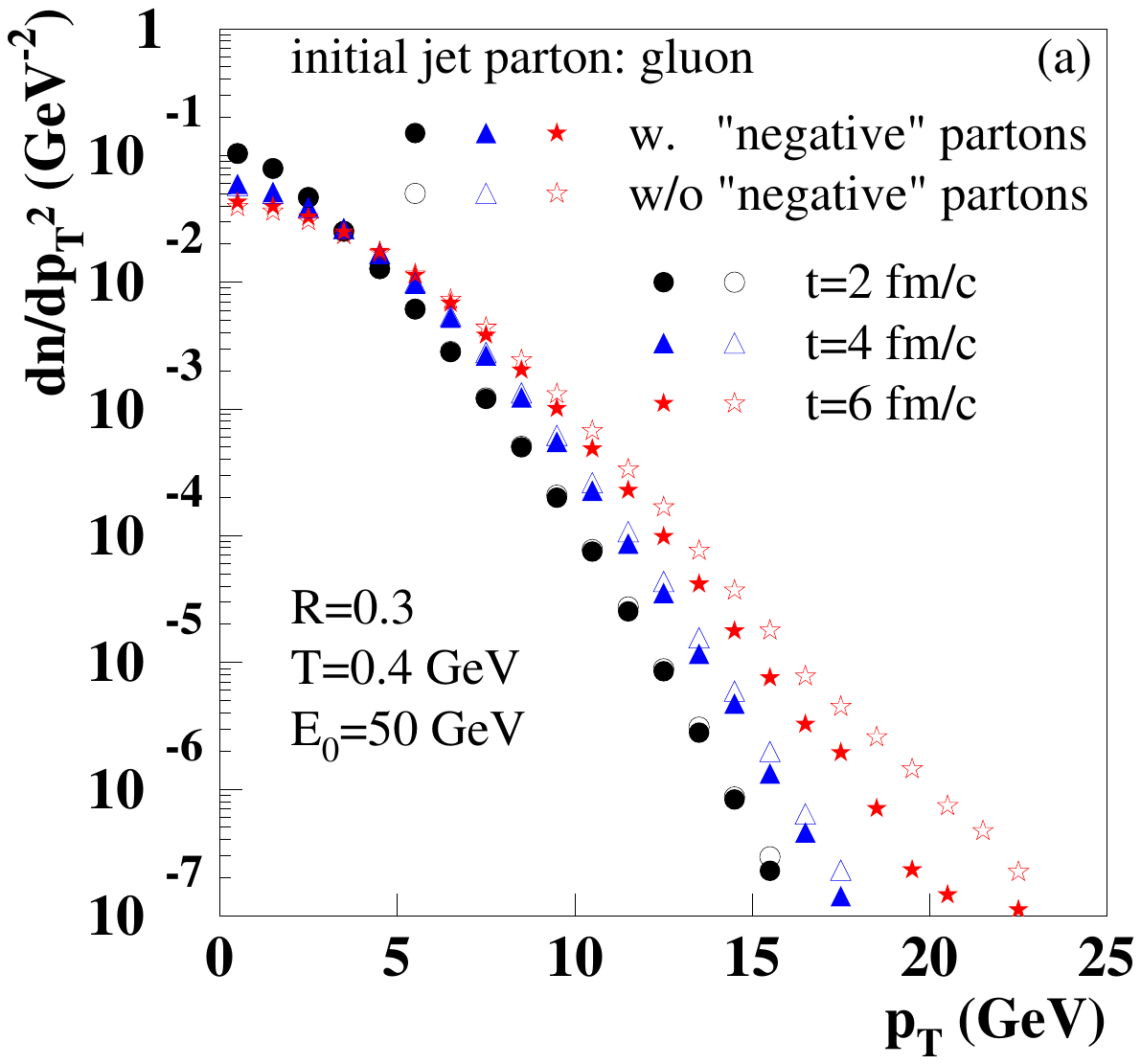}\\
\includegraphics[width=8.5cm,bb=15 150 585 687]{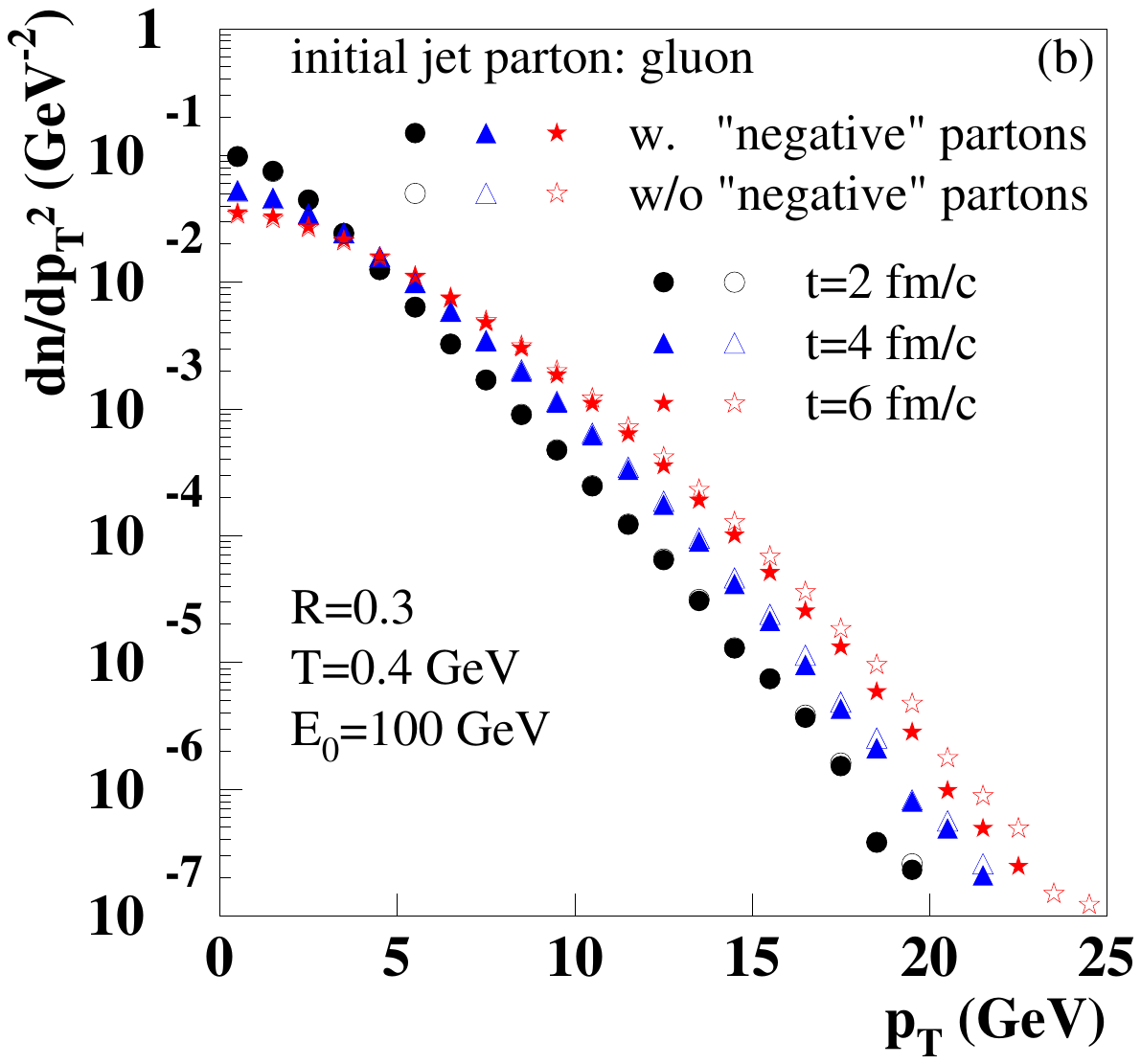}\\
\caption{(Color online) Jet transverse momentum distributions at different times during the propagation of an initial gluon with energy (a) $E_0=50$ GeV and (b) $E_0=100$ GeV in a uniform QGP medium at a temperature $T=400$ MeV. Jets are reconstructed with all partons (leading and jet-induced medium partons) (solid symbols) or without ``negative" partons (open symbols).}
\label{fig-jet-pt}
\end{figure}

During the propagation of an energetic parton, the leading parton experiences transverse momentum broadening through interaction with medium partons. The thermal recoil  parton in the same scattering should also carry the same amount of transverse momentum as the leading parton but in the opposite direction. If a reconstructed jet contains both the leading parton and the thermal recoil   partons (minus the ``negative" partons), its total transverse momentum should not be influenced by the scattering. However, reconstructed jets do not contain all jet-induced medium partons because of the finite jet-cone size $R$. Large-angle scatterings and diffusion of jet-induced medium partons should lead to a reduction of the jet energy,  as shown in the above, as well as a net transverse momentum with respect to the initial parton direction. Shown in Fig.~\ref{fig-jet-pt} are the transverse momentum distributions of reconstructed jets with respect to the initial direction of a propagating gluon with energy $E_0=100$ GeV in a uniform medium at a constant temperature $T=400$ MeV.  There is an apparent transverse momentum broadening at later times during jet transport, similar to the broadening of leading partons. There is also a power law tail in the transverse momentum distribution. This is caused mainly by a single large angle parton-medium scattering during the jet propagation.

\subsection{Medium modification of jet structures}

\begin{figure}
\includegraphics[width=8.5cm,bb=15 150 585 687]{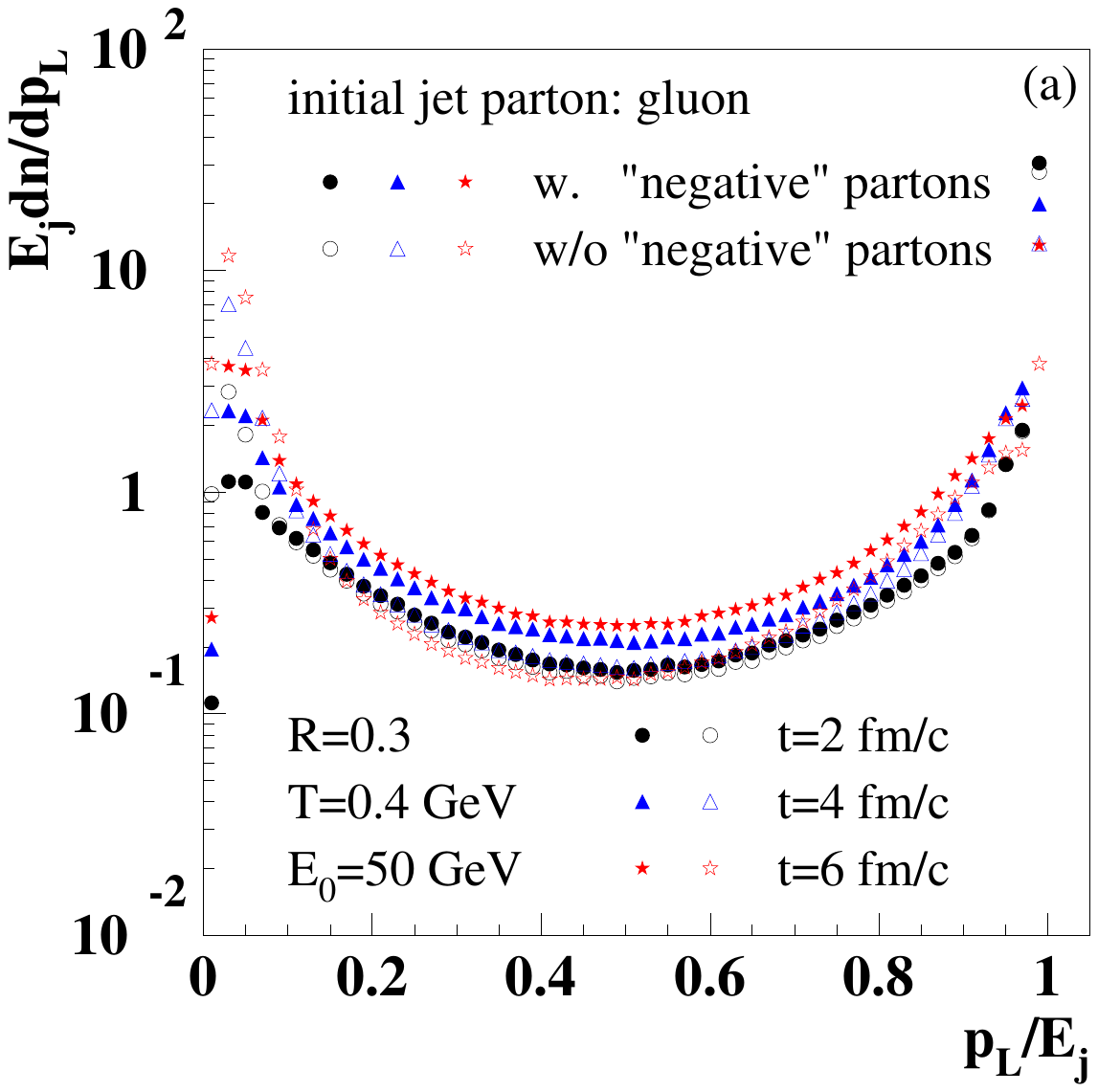}\\
\includegraphics[width=8.5cm,bb=15 150 585 687]{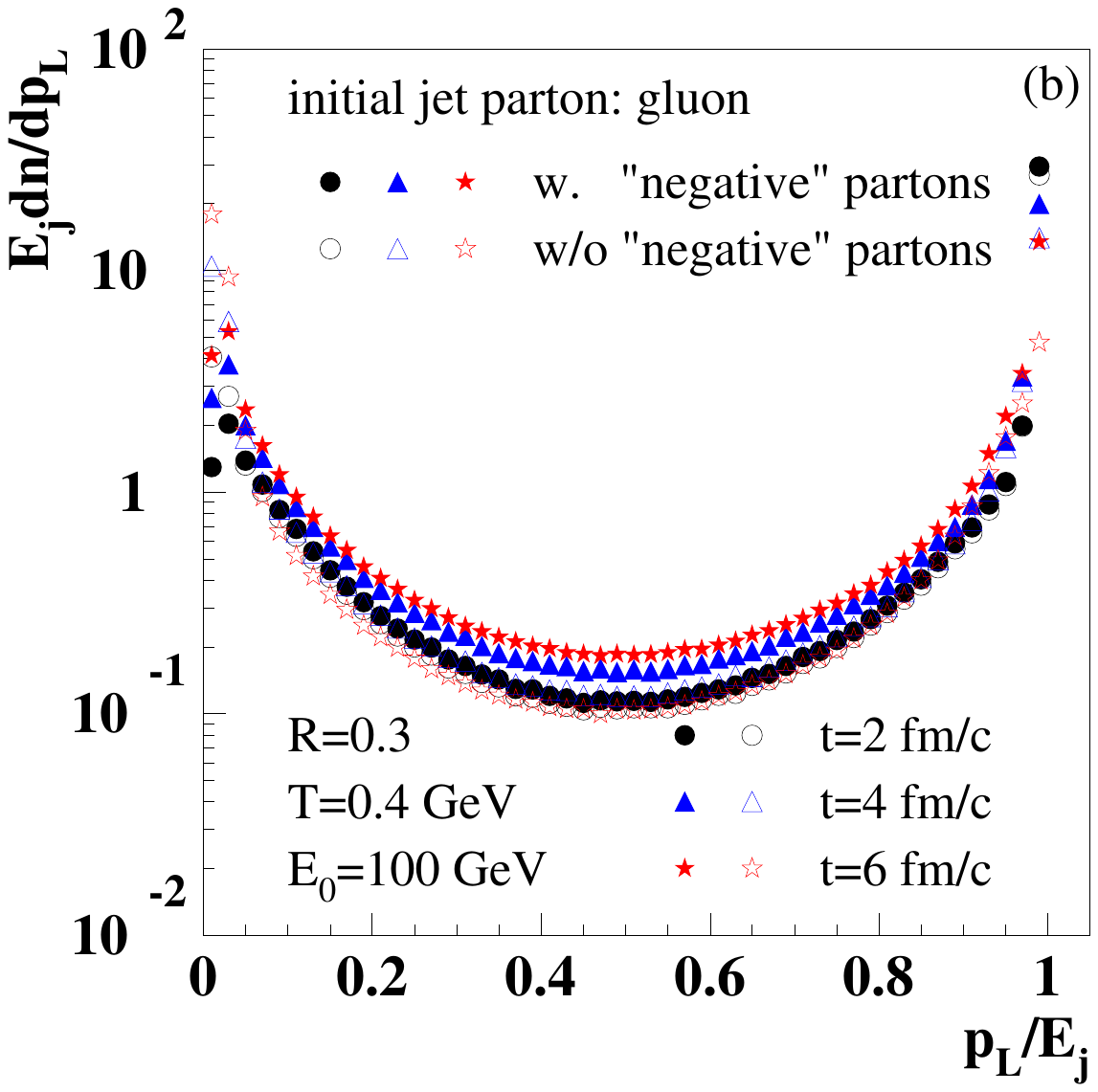}\\
\caption{(Color online) Parton distributions in the momentum fraction $z_J=p_L/E_{\rm jet}$ within the jet cone $R=0.3$ at different times for a propagating gluon with initial energy (a) $E_0=50$ and (b) 100 GeV in a uniform QGP medium at a temperature $T=400$ MeV. Solid symbols represent all partons including leading, recoil, and ``negative" partons while the open symbols show results without ``negative" partons.}
\label{jet-ff}
\end{figure}

\begin{figure}
\includegraphics[width=8.5cm,bb=15 150 585 687]{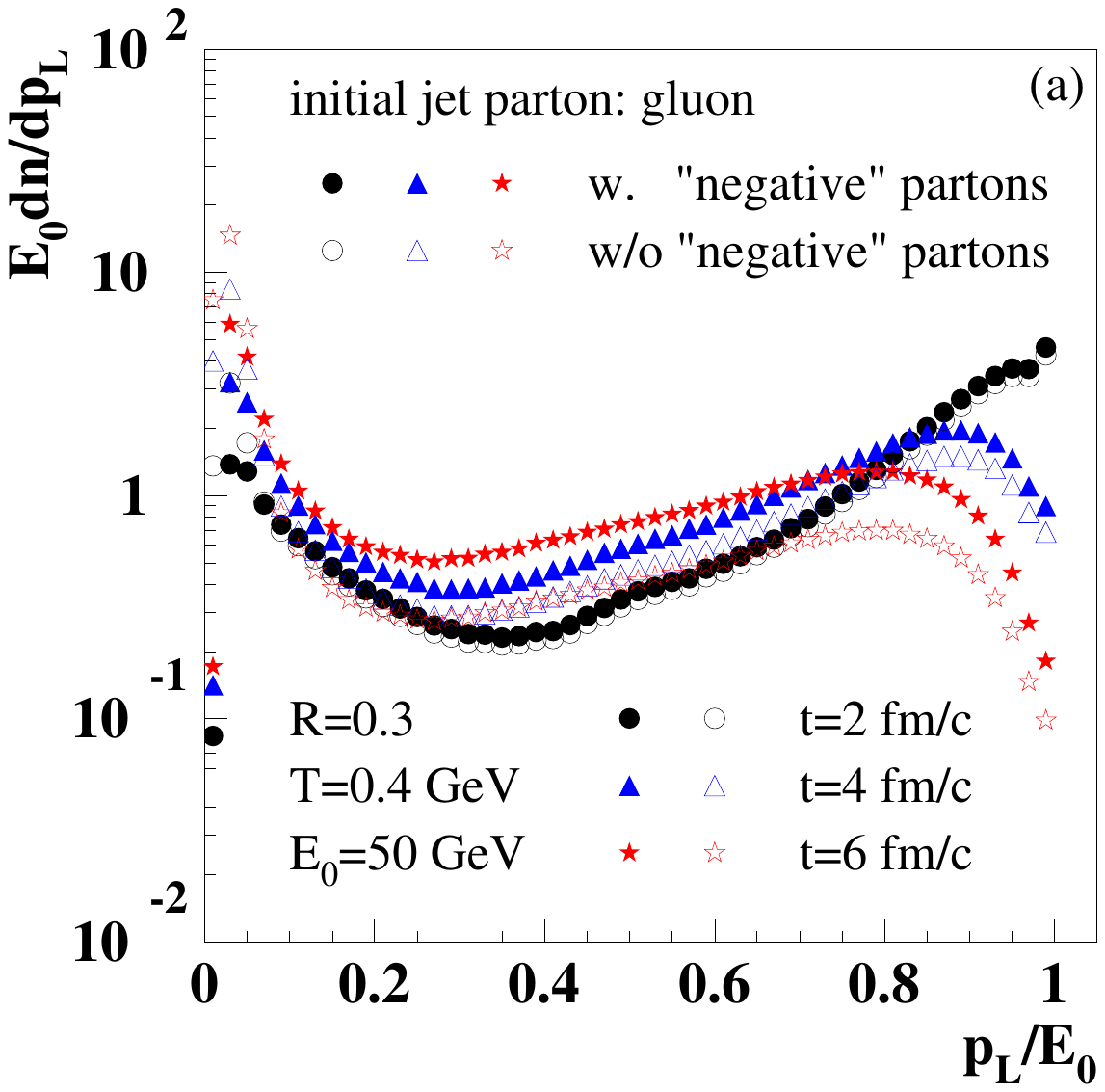}\\
\includegraphics[width=8.5cm,bb=15 150 585 687]{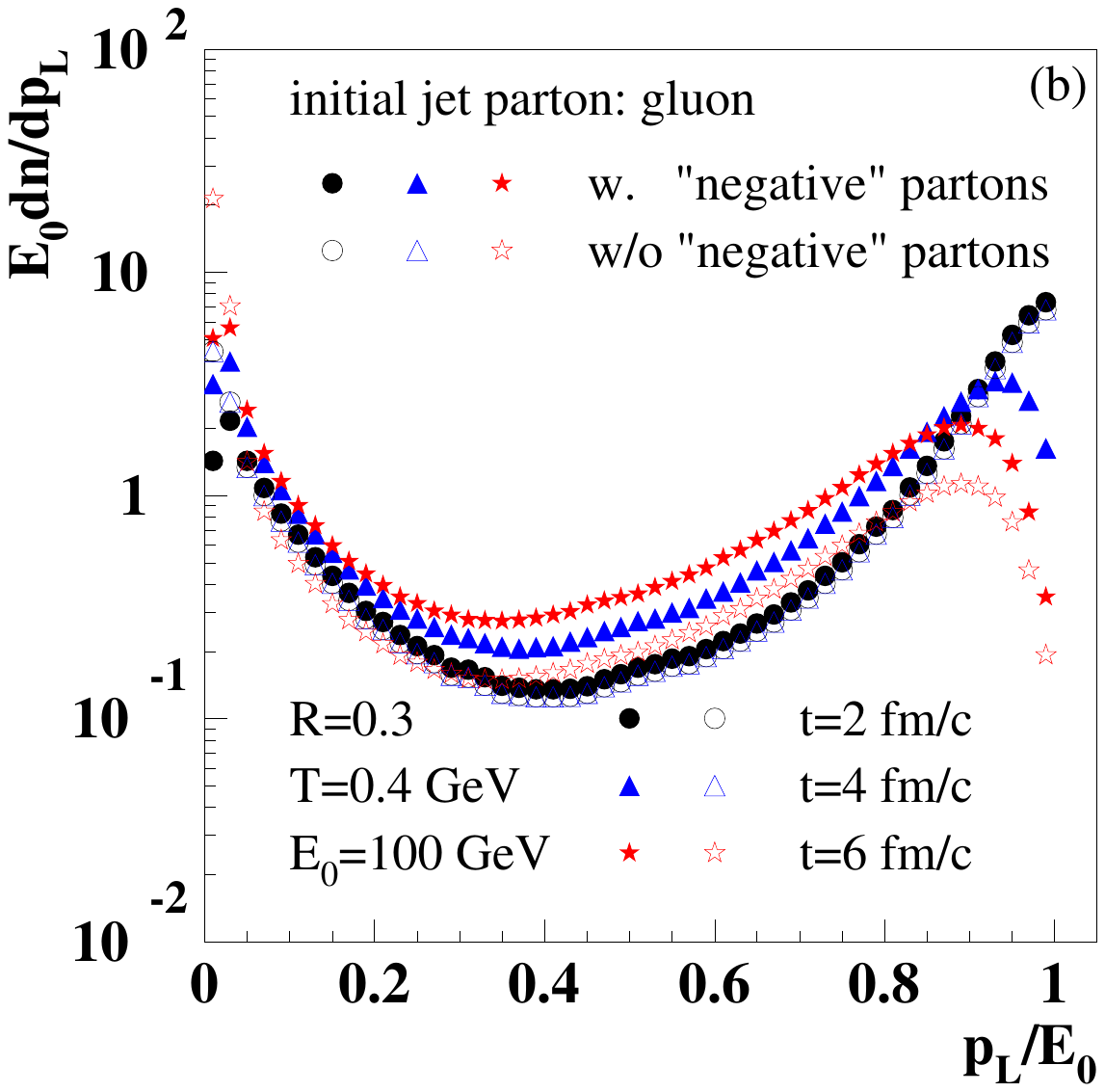}\\
\caption{(Color online) Parton distributions in the momentum fraction $z_0=p_L/E_0$ within the jet cone $R=0.3$ at different times for a propagating gluon with initial energy (a) $E_0=50$ and (b) 100 GeV in a uniform QGP medium at a temperature $T=400$ MeV. Solid symbols represent all partons including leading, recoil, and ``negative" partons while the open symbols show results without ``negative" partons.}
\label{jet-ff0}
\end{figure}

Reconstructed jets in LBT Monte Carlo simulations in this study contain both leading partons and jet-induced medium partons inside the jet cone.  To examine the composition of reconstructed jets and their evolution with time, we show parton distributions within reconstructed jets as functions of the longitudinal momentum fraction $z_J=p_L/E_{\rm jet}$ in Fig.~\ref{jet-ff} and $z_0=p_L/E_0$ in Fig.~\ref{jet-ff0}, where $p_L$ is a parton's longitudinal momentum along the direction of a reconstructed jet with energy $E_{\rm jet}$.  Since we start with a single energetic parton, the initial parton distribution function should be a $\delta$ function at $z_J=z_0=1$. Jet-induced medium partons from parton-medium interaction and leading partons with reduced energy populate the distribution at $z_J<1$ or $z_0<1$ after the jet transport begins. Further scatterings of both leading partons and jet-induced medium partons lead to the enhancement of soft partons at $z_J\ll 1$ or $z_0\ll 1$ as shown in Figs.~\ref{jet-ff} and \ref{jet-ff0}. Since the definition of momentum fraction $z_J$ is normalized by the reconstructed  jet energy $E_{\rm jet}$ which is still dominated by the energetic leading parton, the parton distribution in $z_J$ continues to have a peak at $z_J\sim 1$. The parton distribution in $z_0$ which is normalized by the initial parton energy $E_0$ is, however, suppressed at large $z_0\sim 1$ and its initial peak at large $z_0\sim 1$ disappears at later times as the leading parton continues to lose energy during its propagation through the medium. Similar behavior was reported in the study of $\gamma$-triggered jets in high-energy heavy-ion collisions \cite{Wang:2013cia}. The modified fragmentation functions in terms of momentum fraction of the original jet energy (or the $\gamma$'s energy) are much more sensitive to parton energy loss than that in terms of the momentum fraction of the reconstructed jet energy. Since ``negative" partons are subtracted from both the parton distribution and the reconstructed jet energy (reducing $E_{\rm jet}$), they will reduce the soft parton distributions at small $z_J\ll1$ or $z_0\ll 1$ and enhance the parton distribution at large $z_J\sim 1$.

\begin{figure}
\includegraphics[width=8.5cm,bb=15 150 585 687]{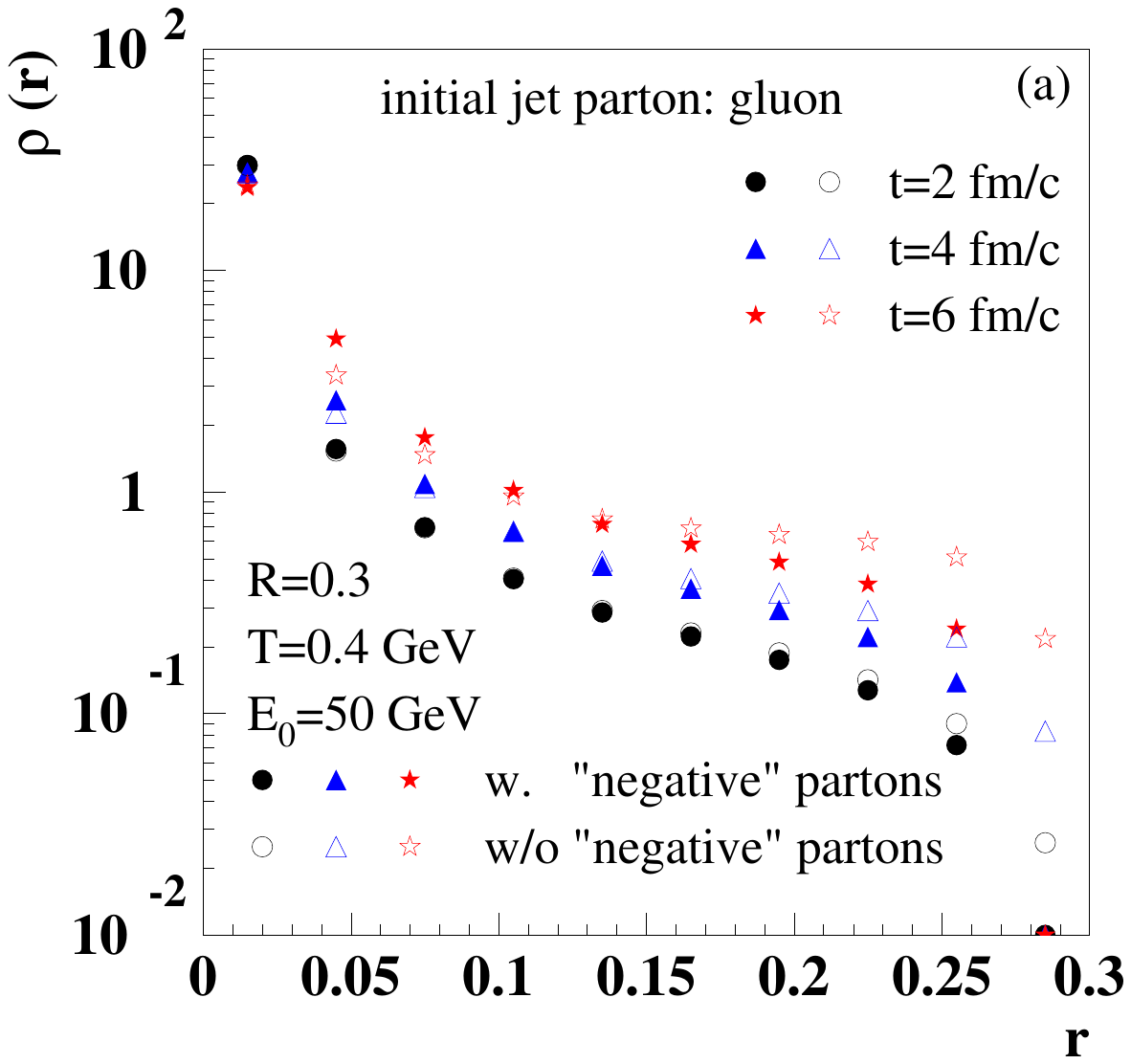}\\
\includegraphics[width=8.5cm,bb=15 150 585 687]{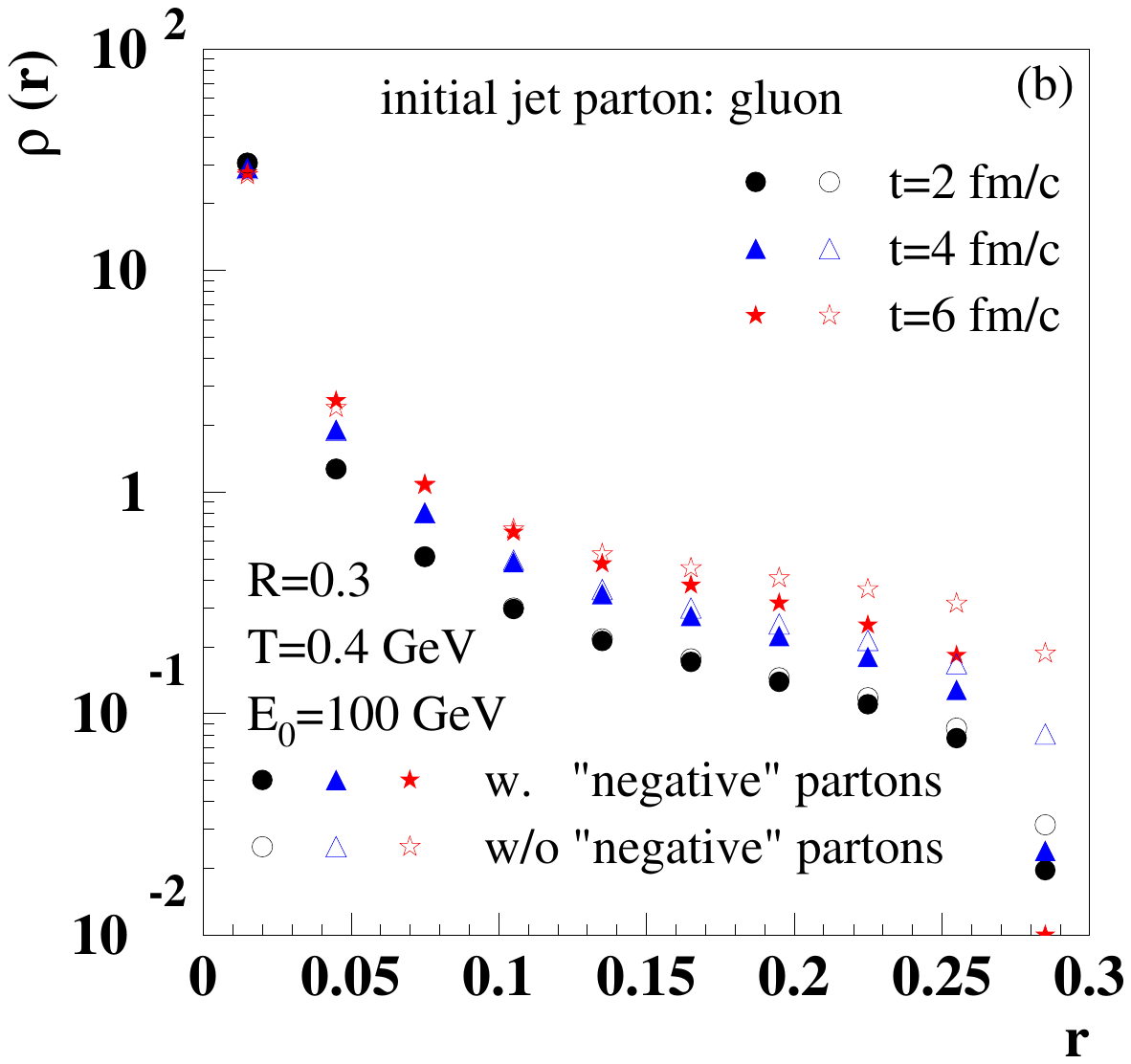}\\
\caption{(Color online) Jet transverse profile of the leading jet by a propagating gluon with initial energy (a) $E_0=50$ and (b) 100 GeV at different times in a static and homogeneous medium at temperature \(T=400\) MeV. Results that include all (leading+positive+negative) partons are presented in solid symbols while results without ``negative" partons are presented in open symbols. }
\label{jet-profile}
\end{figure}

To study the evolution of the jet transverse structure, we show in Fig.~\ref{jet-profile} the transverse profile of reconstructed jets at different times for a propagating gluon with an initial energy $E_0=50$ GeV (upper panel) and $E_0=100$ GeV (lower panel) in a uniform QGP medium at a constant temperature $T=400$ MeV. The jet transverse profile is defined as the average fraction of jet energy inside an annulus in the $\eta-\phi$ plane,
\begin{equation}
\rho(r)=\frac{1}{\Delta r}\frac{1}{N^\text{jet}}\sum_\text{jets}\frac{E(r-\Delta r/2,r+\Delta r/2)}{E(0,R)},
\end{equation}
as a function of $r=\sqrt{(\eta-\eta_J)^2+(\phi-\phi_J)^2}$, where $E(r_1,r_2)$ is the summed energy of all partons in the annulus between radius $r_1$ and $r_2$ inside the jet cone. When a high energy parton propagates through the plasma, it loses energy to the medium. The lost energy is carried away by thermal recoil  partons whose further diffusion through multiple scatterings will transport some of the lost energy outside the jet cone leading to the jet energy loss and broadening of the transverse profile toward the edge of the jet cone, as seen in Fig.~\ref{jet-profile}.  The ``negative" partons are mostly soft partons at large angles away from the initial parton's direction. When their energy is subtracted from the reconstructed jet, the broadening of the transverse profile will be reduced, especially near the edge of the jet cone.

\section{Summary and Discussions}

In this paper we have described in detail the basic elements of the LBT model for the study of jet transport in a QGP medium with the complete set of elastic $2\rightarrow 2$ scattering processes. We verified the Monte Carlo code in the LBT model by direct comparisons to semi-analytic results of total and differential scattering rates via numerical integrations. We calculated the transverse momentum broadening and elastic energy loss of the leading parton along the path of the parton propagation. Within the pQCD approach to each elastic scattering between the leading and thermal medium parton, both the jet transport coefficient $\hat q$ and the energy loss per mean-free-path have a logarithmic dependence on the parton energy.  Such an energy dependence leads to some nontrivial time or distance dependence  of the average transverse momentum broadening and energy loss per unit distance due to the decrease of the leading partons' energy over time. This dependence is more significant when the total energy loss becomes comparable to the initial parton's energy.

We have also illustrated the dissipation of the energy and momentum lost by a propagating parton in the medium via transport of jet-induced medium partons (thermal recoil  partons and ``negative" partons from the back-reaction) in the LBT model. The transport of these jet-induced medium partons within the LBT model effectively forms a supersonic shock wave and a diffusion wake behind the leading parton. The supersonic wave has a diffused Mach-cone shape because of the transverse momentum broadening of the leading parton along its path. The energy spectra of jet-induced medium partons in the supersonic wave are shown to resemble that of a thermal distribution at later times of the jet propagation.

Using a modified version of the FASTJET jet-finding package with the anti-$k_T$ algorithm \cite{Cacciari:2011ma} in which the energy of ``negative" partons from the back-reaction is subtracted, we show the effect of jet-induced medium partons and their dissipation in medium on reconstructed jets. Transport of these jet-induced medium partons outside the jet cone constitutes a significant reduction of jet energy loss which should be taken into account for any realistic study of jet suppression in high-energy heavy-ion collisions. Their inclusion in the jet reconstruction also influences the transverse momentum broadening, parton distributions (fragmentation functions) and transverse profiles of reconstructed jets.

Transverse momentum distributions of both leading partons and reconstructed jets are shown to have a power law tail underneath Gaussian distributions from multiple scattering. These power law tails come from a single large angle parton-medium scattering during the jet propagation. Study of these power law tails in heavy-ion collisions can shed light on the microscopic nature of the QGP medium at different scales.

For a realistic description of jet transport in the QGP medium in high-energy heavy-ion collisions, we will have to implement inelastic processes such as gluon bremsstrahlung induced by multiple parton-medium scattering \cite{Wang:2013cia,Wang:2014hla} and couple LBT Monte Carlo simulations with bulk medium evolutions from 3+1D hydrodynamic model calculations, which should be constrained by bulk hadron spectra from existing experimental data. These parts of the LBT model and the phenomenological study of jet suppression in heavy-ion collisions will be discussed in subsequent publications.

\begin{acknowledgments}
We thank Wei Chen for helpful discussions, in particular on the elliptic paraboloid shape of the shock  wave induced by a propagating parton.
This work is supported by China MOST under Grant No. 2014DFG02050, the NSFC under Grant No. 11221504, the Major State Basic Research Development Program in China (Grant No. 2014CB845404), by the Director, Office of Energy
Research, Office of High Energy and Nuclear Physics, Division of Nuclear Physics of the U.S. Department of
Energy under Contract No. DE-AC02-05CH11231 and within the framework of the JET Collaboration. Y.-Z. is also supported by European Research Council Grant No. HotLHC ERC-2011-StG-279579.

\end{acknowledgments}

\end{document}